\documentclass[printer]{aa}

\usepackage{graphicx}
\usepackage{amsmath}
\usepackage{txfonts}
\newcommand{\ee}[1]{#1}   %%does nothing

\newcommand{\out}[1]{}  %%take out (hide) text within command domain

\newcommand{\mapc}{\hbox{{\sc mappings i}c}}
\newcommand{\map}{\hbox{{\sc mappings i}}}
\newcommand{\mape}{\hbox{{\sc mappings i}e}}

\def\ion#1#2{#1$\;${\small\rm\@Roman{#2}}\relax}

\newcommand{\up}{\hbox{$U$}}
\newcommand{\upa}{\hbox{$U_A$}}
\newcommand{\upb}{\hbox{$U_B$}}
\newcommand{\cms}{\hbox{\,${\rm cm^{-2}}$}}
\newcommand{\cmc}{\hbox{\,${\rm cm^{-3}}$}}
\newcommand{\etal}{\hbox{et~al.}}
\newcommand{\msup}{\hbox{M$_{sup}$}}
\newcommand{\msol}{\hbox{\,${\rm M_{\sun}}$}}

\newcommand{\kms}{\hbox{${\rm km\,s^{-1}}$}}

\newcommand{\dex}{\hbox{\,{dex}\,}\/}

\newcommand{\av}{\hbox{A$_{V}$}}

\newcommand{\hb}{\hbox{H$\beta$}}

\newcommand{\hi}{\hbox{H{\sc i}}}
\newcommand{\hii}{\hbox{H{\sc ii}}}
\newcommand{\heii}{\hbox{He\,{\sc ii}}}

\newcommand{\nii}{\hbox{[N\,{\sc ii}]}}

\newcommand{\niitw}{\hbox{[N\,{\sc ii}]$\lambda $5755}}
\newcommand{\oiiw}{\hbox{[O\,{\sc ii}]$\lambda\lambda $3727,29}}
\newcommand{\oii}{\hbox{[O\,{\sc ii}]}}
\newcommand{\oi}{\hbox{[O\,{\sc i}]}}

\newcommand{\sii}{\hbox{[S\,{\sc ii}]}}
\newcommand{\siiw}{\hbox{[S\,{\sc ii}]$\lambda\lambda $6716,31}}
\newcommand{\oiii}{\hbox{[O\,{\sc iii}]}}
\newcommand{\ariii}{\hbox{[Ar\,{\sc iii}]}}

\newcommand{\siii}{\hbox{[S\,{\sc iii}]}}

\newcommand{\siiitw}{\hbox{[S\,{\sc iii}]$\lambda $6312}}
\newcommand{\oiiiw}{\hbox{[O\,{\sc iii}]$\lambda $5007}}
\newcommand{\oiiitw}{\hbox{[O\,{\sc iii}]$\lambda $4363}}

\newcommand{\tsiii}{\hbox{$T_{S{III}}$}}
\newcommand{\toiii}{\hbox{$T_{O{III}}$}}

\newcommand{\nh}{\hbox{$n_{\rm H}$}}

\newcommand{\Nhptot}{\hbox{$N_{\rm H+}^{tot}$}}
\newcommand{\Nhpsho}{\hbox{$N_{\rm H+}^{sho}$}}

\newcommand{\npre}{\hbox{$n_{\rm pre}$}}
\newcommand{\npost}{\hbox{$n_{\rm post}$}}
\newcommand{\vs}{\hbox{$V_{s}$}}
\newcommand{\sed}{\hbox{{\sc sed}}}
\newcommand{\teq}{\hbox{$T_{eq}$}}
\newcommand{\tmean}{\hbox{$T_{\!o}$}}
\newcommand{\trec}{\hbox{$T_{rec}$}}
\newcommand{\treco}{\hbox{$T_{rec}^{O++}$}}
\newcommand{\trecx}{\hbox{$T_{rec}^{X+i}$}}
\newcommand{\trech}{\hbox{$T_{rec}^{H+}$}}
\newcommand{\treche}{\hbox{$T_{rec}^{He+}$}}

\newcommand{\teff}{\hbox{$T_{eff}$}}
\newcommand{\tpre}{\hbox{$T_{pre}$}}
\newcommand{\tpost}{\hbox{$T_{post}$}}
\newcommand{\tsq}{\hbox{$t^2$}}
\newcommand{\zsol}{\hbox{$Z_{\sun}$}}

\begin{document}

%%%%%  TITLE & AUTHOR/ADDRESS  PAGE     %%%%%%%%%%%%%%%%%%%%%%%%%%%%%%%%%%%%%

   \title{Discrepancies between the \oiii\ and \siii\  \\temperatures in \hii\ regions}

   \author{Luc Binette\inst{1,2}
            \and
          Roy Matadamas\inst{2}
         \and
          Guillermo F. H\"agele\inst{3,4,5}
          \and
          David C. Nicholls\inst{6}
          \and
          Gladis Magris C.\inst{7}
          \and
          \\Mar\'{i}a de los \'Angeles Pe\~na-Guerrero\inst{8,2}
          \and
          Christophe Morisset\inst{2}
          %\fnmsep\thanks{Just to show the usage
          %of the elements in the author field}
          \and
          Ary Rodr\'{i}guez-Gonz\'alez\inst{9}
          }

   \institute{D\'{e}partement de Physique, de G\'{e}nie Physique et
        d'Optique, Universit\'{e} Laval, Qu\'{e}bec, QC, G1V\,0A6, Canada
        \and
   Instituto de Astronom\'{i}a, Universidad Nacional Aut\'onoma de M\'exico, D.F., Mexico
   %\thanks{RM email: rlopez@astro.unam.mx}
         \and
             Consejo Nacional de Investigaciones Cient\'{i}ficas y T\'ecnicas (CONICET), Argentina
%             \email{guille.hagele@uam.es}
%             \thanks{}
         \and
             Facultad de Ciencias Astron\'omicas y Geof\'{i}sicas, Universidad Nacional de La Plata, Paseo del Bosque s/n, 1900 La Plata, Argentina
%             \email{guille.hagele@uam.es}
%             \thanks{}
         \and
             Departamento de F\'{i}sica Te\'{o}rica, C-XI, Universidad Aut\'{o}noma de Madrid, 28049 Madrid, Spain
%             \email{guille.hagele@uam.es}
%             \thanks{}
         \and
       Research School of Astronomy and Astrophysics, Australian National University, Cotter Rd., Weston ACT 2611, Australia
         \and
             Centro de Investigaciones de Astronom\'{i}a, Av. Alberto Carnevalli, M\'erida, Venezuela
%             \email{magris@cida.ve}
%             \thanks{}
        \and
             Space Telescope Science Institute, Baltimore, Maryland 21218, USA
        \and
            Instituto de Ciencias Nucleares, Universidad Nacional Aut\'onoma de M\'exico, Ap. 70-543, 04510 D.F., Mexico
%             \email{guerrero@astro.unam.mx}
%             \thanks{}
              }

%  \date{Received September 15, 1996; accepted March 16, 1997}

 %%%%%%%%%%%%%%%%%%%%%%%%%%%%% ABSTRACT  A&A STYLE  %%%%%%%%%%%%%%%%%%%%%%%%%%%%%%%%%%%%%%%%%%%%%%%

 \abstract{Analysis of published \oiii\  and \siii\ temperatures measurements of emission line objects consisting of \hii\ galaxies, giant extragalactic \hii\ regions, Galactic \hii\ regions, and \hii\ regions from the Magellanic Clouds reveal  that  the \oiii\ temperatures are higher than the corresponding values from \siii\ in most objects with gas metallicities in excess of 0.2 solar.  For the coolest nebulae (the highest metallicities), the \oiii\ temperature excess can reach  $\sim 3\,000$\,K.}
 {We look for an explanation for these temperature differences and explore the parameter space of models with the aim of reproducing the observed trend of $\toiii > \tsiii$ in \hii\ regions with temperatures below 14\,000\,K.}
 {Using standard photoionization models, we varied the ionization parameter, the hardness of the ionizing continuum, and the gas metallicities in order to characterize how models behave with respect to the observations. We introduced temperature inhomogeneities and varied their mean squared amplitude \tsq. We explored the possibility of inhomogeneities in abundances by combining two models of widely different metallicity. We calculated models that consider the possibility of a non-Maxwell-Boltzmann energy distribution (a $\kappa$--distribution) for the electron energies. We also considered shock heating within the photoionized nebula.}
 {Simple photoionization calculations yield nearly equal \oiii\ and \siii\ temperatures in the domain of interest. Hence these models fail to reproduce the \oiii\ temperature excess. Models that consider temperature inhomogeneities, as measured by the mean squared amplitude \tsq, also fail in the regime where $\toiii < 14\,000$\,K.  Three options remain that can reproduce the observed excess in \toiii\ temperatures: (1) large metallicity inhomogeneities in the nebula, a (2) $\kappa$--distribution for the electron energies, and (3) shock waves that propagate in the photoionized plasma at velocities $\sim 60$\,\kms. }
 {The observed nebular temperatures are not reproduced by varying the input parameters in the pure photoionization case nor by assuming local temperature inhomogeneities. We find that (1) metallicity inhomogeneities of the nebular gas, (2) shock waves of velocities $\la 60$\,\kms\ propagating in a photoionized plasma, and (3) an electron energy distribution given by a $\kappa$--distribution are successful  in reproducing the observed excess in the \oiii\ temperatures. However, shock models  require proper 3D hydrodynamical simulations to become a fully developed alternative while models with metallicity inhomogeneities appear to fail in metal-poor nebulae, since they result in $\treco \ga \toiii$.}


   \authorrunning{L. Binette et\,al.}
   \titlerunning{TEMPERATURE DISCREPANCIES IN \hii\ REGIONS}

   \maketitle
%________________________________________________________________

%%%%%%%%%%%%%%%%%%%%%%%%%%%%%%%%%%%%%%%%%%%%%%%%%%%%%%%%%%%%%%%%%%%%%%%%%%
%% And here starts the text....    INTRODUCTION

\section{Introduction} \label{sec:intro}

%Many features of star formation can be studied knowing the conditions of the ionized gas that surround young massive stars formed in the core of  giant molecular clouds.

The emission and absorbtion lines of star-forming regions can inform us about the physical conditions of the gaseous media, such as abundances, temperature, and ionization degree. They also provide us with estimates of the ages, masses, and composition of the stellar population and the properties of the ionizing stellar clusters (see e.g.\ H\"agele \etal\ 2011, hereafter H11; and references therein). The electron temperature is a property crucial for interpreting emission line intensities.  When we investigated the \siii\  temperatures in the literature, the study of H\"agele \etal\ (2006, hereafter H06), who compared the  temperatures inferred from collisionally excited lines  of \siii\ with those of \oiii, caught our attention. Reproducing these with photoionization models was set by some of us (RM and LB) as a potentially instructive exploratory exercise that turned out more challenging as we proceeded with our analysis. Our paper reports on this exploration and is a work in progress, mainly because the uncertainties in the data are substantial, as discussed in Sect.\,\ref{sec:dat}. In essence, in nebulae where measurement errors are the  smallest, we find that the \toiii\ temperatures exceed the values derived for \tsiii\ by up to $\sim 3\,000\,$K, a trend that we were \emph{not} able to reproduce using simple photoionization calculations. We consider it remarkable (yet with insight predictable) that the relationship between these two temperatures is quite insensitive to strong changes in the input parameters of photoionization models. We subsequently explored models with (1) metallicity inhomogeneities, (2) temperature inhomogeneities, (3) non-Maxwell-Boltzmann energy distributions (hereafter a $\kappa$--distribution), and completed our investigation with models that (4) combine photoionization with shock heating.  We hope that the current work may motivate observers to undertake observational studies that are able to substantially reduce errors in the measurements of the nebular and auroral \siii\ lines.

Considering the extra difficulties in measuring the \siii\ lines in the infrared and the auroral counterpart at 6312\AA, what alternative species can be used to infer the electron temperature from an auroral-to-nebular line ratio? There are the two well-known  \nii\ or \oii\ temperature diagnostics, but they carry their own set of uncertainties, which renders an interpretation of these diagnostics ambiguous and more model dependent. For instance, the electron temperatures deduced from the \nii\ $5755\AA/(6548\AA+6583\AA)$ ratio are frequently higher than those of the corresponding \oiii\ lines (e.g. Tsamis \etal\ 2003).  According to Stasinska (1980) and Copetti (2006), the increase in temperature outward as a function of nebular radius is likely caused by a combination of hardening of the radiation field with increasing optical depth, and stronger cooling from fine-structure lines of \oiii\ in the inner parts of the nebulae where O$^{+2}$ dominates (see Stasinska 1980).
This expectation is not fully confirmed, however, due to the significant contribution of recombination to the excitation of the auroral \niitw\ line (Rubin 1986; Tsamis \etal\ 2003) coupled with the potential presence of high-density inclusions in nebulae (Viegas \& Clegg 1994), all of which indicates that \nii\ fails to qualify as a suitable alternative.  The \oii\ nebular-to-auroral line ratio is even more susceptible to contribution from recombination and to the potential presence of high-density inclusions in nebulae (Viegas \& Clegg 1994), which may cause deceptively high temperatures to be derived from the \oii\ $3727\AA/(7320\AA+7330\AA)$ ratio. Furthermore, this line ratio is more sensitive to variations in the extinction curve ($R_V$) or to second-order effects from an inhomogeneous dust distribution, which standard reddening corrections cannot correct for. In comparison to N$^+$ and O$^+$, the specie S$^{+2}$ presents the clear advantage of strongly overlapping O$^{+2}$ in the higher excitation region of a nebula. An appealing alternative might be the \ariii\ auroral-to-nebular line ratio  $5192\AA/(7136\AA+7751\AA)$ (Keenan \etal\ 1988; Copetti 2006), but there are very little data available for these lines owing to their intrinsic weakness.

Studying physical properties of emission line objects at different redshifts with accurate and reliable methods is a fundamental issue when one aims to detect possible evolutionary effects such as systematic differences in the chemical composition of these objects.  We discuss in Sect.\,\ref{sec:res}  that the observation that the \toiii\ temperatures are higher than \tsiii\ may be connected to gaseous nebula phenomena, which are not yet fully understood. One of these phenomena is the abundance discrepancy factor (ADF), which is the ratio of the abundance derived for a heavy-element ion from its optical recombination lines (ORLs) to the abundance derived for the same ion from its ultraviolet, optical, or infrared collisionally excited lines (CELs) (Tsamis 2004). In \hii\ regions, the ADF is about 2 and has been the subject of various interpretations (Peimbert \etal\ 1995; Esteban \etal\ 2002; Tsamis \& P\'equignot 2005; P\'erez-Montero \& D\'iaz 2007;  Tsamis \etal\ 2003, 2011; Nicholls \etal\ 2012; L\'opez-S\'anchez et\,al. 2012, and references therein).

For the present study we carried out several models using the multipurpose photoionization-shock code \mape\ that allows us to explore five options: (1) simple photoionization models where we varied the main input parameters, (2) models that consider the effect of temperature inhomogeneities\footnote{Following Pe\~na-Guerrero et\,al. (2012a), we prefer the terminology `temperature inhomogeneities' over that of `temperature fluctuations' to avoid confusions with temporal fluctuations (e.g. Binette \etal\ 2003).}, (3) a combination of models of widely different metallicities,  (4) models that consider $\kappa$--distributions of electron energies (non-Maxwell-Boltzmann energy distributions), and (5) models that combine shock excitation with photoionization. The article is structured as follows: Section\,2 contains the data analysis, our modeling strategy is described in Section\,3, and the comparison of models with the data is discussed in Section\,4. Brief conclusions are presented in Section\,5.

%% THE DATA BASE

\section{The data base}\label{sec:dat}

\begin{figure}
\begin{center} \leavevmode
\includegraphics[width=0.49\textwidth]{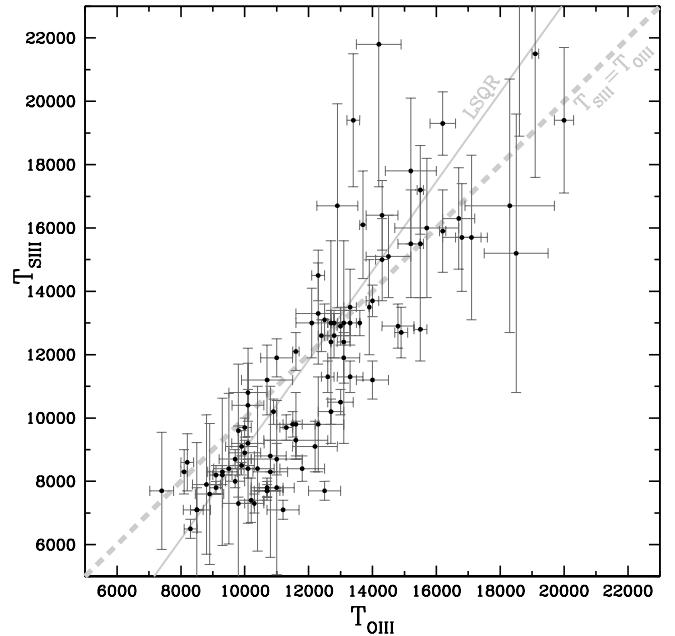}
\caption{Relation between \tsiii\ and \toiii\ temperatures inferred from the \siii\ and \oiii\ auroral-to-nebular line ratios for the objects studied by PM06, including the three \hii\ galaxies later observed by H06. The light gray dashed line depicts the locus of equal temperatures ($\tsiii=\toiii$), while the solid line indicates the error-weighted least-squares fit to the data (see Sect\,\ref{sec:dat}).}
\label{fig:plterr}
\end{center}
\end{figure}

%The \hii\ galaxies consist of low-mass irregular galaxies revealing at least one recent episode of violent star formation (Melnick, Terlevich \& Eggleton, 1985; Melnick, Terlevich \& Moles 1985). They are centered within a few parsecs of their cores. The ionizing flux that originates from the young massive stars dominates the light of this subclass of Blue Compact Dwarf galaxies (BCDs), which show emission line spectra very similar to those of GEHRs (Sargent and Searle, 1970; French 1980). By using the same measurement techniques as for \hii\ regions, it is possible to derive the temperatures, densities, and chemical composition of the interstellar gas in this type of generally metal-deficient galaxies (Terlevich et\,al. 1991; Kunth \& Ostlin, 2000; Hoyos \& D\'iaz, 2006).

The analysis carried out in this paper relies on two temperature-sensitive line ratios that involve auroral and nebular line transitions of the same ion. These are \oiii\ 4363\AA/(4959\AA+5007\AA) and \siii\ 6312\AA/(9069\AA+9532\AA). The inferred temperatures assuming the low-density limit hereafter are labeled \toiii\ and  \tsiii.  Our sample corresponds to a compilation of data (99 objects) initially published by P\'erez-Montero \etal\ (2006: hereafter PM06) and subsequently augmented by three \hii\ galaxies observed by H06. In total, the sample consists of the spectra of 42 \hii\ galaxies, 34 giant extragalactic \hii\ regions (hereafter GEHRs), 14 Galactic \hii\ regions, and 12 \hii\ regions from the  Magellanic Clouds. Figure~\ref{fig:plterr} shows the temperatures inferred from a total of 102 pairs of \oiii\ and \siii\ line ratio measurements as well as the associated error bars.  For clarity's sake, the error bars are not drawn in the figures. Hereafter, we will use the terms \emph{\hii\ region} or \emph{nebula} as generic expressions representative of objects from the sample.
%interchangeable

As {is evident from the distribution}, a  correlation between the two temperatures is observed in Figure~\ref{fig:plterr}. An error-weighted least-squares fit of the data results in the following linear regression:
\begin{equation}
\tsiii = 1.41\pm{0.0944} \times \toiii - 5090\pm{1305} \; {\rm K},  \label{eq:dat}
\end{equation}
\noindent with a correlation coefficient of 0.85. The light gray dashed line depicts what would be the locus of \emph{equal} temperature values in this plot and in subsequent figures. The correlation is markedly steeper than that inferred by Kennicutt \etal\ (2003) of  $\tsiii = 0.83 \times \toiii - 1700 \,{\rm K}$ from a study of 20\,\hii\ regions in the spiral galaxy M101 (these objects are included in our much larger sample). We note a systematic inequality between the two temperatures in Fig.\,\ref{fig:plterr}. In particular, lower temperature objects $\tsiii < 14\,000$\,K tend to lie below the equal temperature dashed line.  Towards the high temperature end ($\tsiii > 14\,000\,$K), however, the error bars and the overall dispersion of the different measurements become quite large. Due to this and to the absence of any clear trend in the data, our analysis will disregard the high temperature upper right quadrant in Fig.\,\ref{fig:plterr}.

Given the distribution of the error bars and the increasing dispersion of the points toward higher temperatures, it is doubtful that a single linear regression provides an appropriate description of the current data set. We suggest that a bimodal distribution might be a better description of the data: 1) the lower left quadrant, which shows measurements that are correlated and possess smaller error bars, and 2) the upper right quadrant, where no clear trend is apparent and where the error bars are noticeably larger.  Our study and the modeling presented below focus on the temperature behavior within this lower left quadrant and models the temperature departure from the dashed line in Fig.\,\ref{fig:plterr}, that is, the domain  $\toiii > \tsiii$ below $\toiii < 14\,000$\,K. Even if the error bars had been underestimated by PM06, a proportionally higher noise amplitude would not erase the general trend that the \oiii\ temperatures lie above those of \siii.  Could this trend be the result of a bias affecting the weaker lines? Rola \& Pelat (1994)  established that the 'measured' intensities in poor signal-to-noise (S/N) line measurements are strongly biased toward overestimating their intrinsic values.  The errors reported by PM06 for the vast majority of objects are markedly larger for \tsiii\ than for \toiii. Because auroral lines are much weaker and the error in subtracting the underlying  continuum larger, the errors pertaining to the weaker auroral lines are larger. The statistical bias studied by  Rola \& Pelat (1994) would in our case lead to an \emph{overestimate} of \siiitw, hence of \tsiii, which is the opposite of the observed trend.  Telluric absorption of the infrared \siii\ 9069\AA\ and 9532\AA\ lines or contamination of the 6312\AA\ line by atmospheric emission\footnote{This contamination would only take place for certain redshifts or when the spectral resolution is insufficient.}  from nearby \oi\ 6300\AA\ would both contribute to enhancing the inferred \tsiii\ values, in contrast to the observed trend. In our view, the spread in temperatures in the lower left quadrant is genuine, but we cannot rule out that unforeseen systematic errors are contributing to it.

Improving the accuracy of temperature measurements will be required before definitive conclusions can be drawn.  The  infrared \siii\ lines are intrinsically more difficult to measure. The risk of systematic errors caused by atmospheric telluric emission, which increases toward longer wavelengths near 9500\AA\ and beyond, remains a concern.  As databases will grow, a significant improvement in the quality of the data should be the first priority. For instance, it might become preferable to discard data for which the 9532\AA/9069\AA\ ratio does not agree with the theoretical value of 2.46 (from Podobedova \etal\ 2009) within a reasonable S/N margin.

%% THE 4 MODELLING APPROACH USED

\section{Model calculations} \label{sec:cal}
%solar respecto al valor de Asplund et al. (O/H=4.6e-4)

An abundance analysis of the complete sample has already been carried out by H06 and PM06, in which the authors took into account the measured nebular temperatures as well as the intensities of the forbidden lines with respect to the recombination lines. Such an abundance analysis will not be repeated in this paper. Rather, our aim will be to investigate whether one can model  \oiii\ temperatures that are higher than those derived from \siii, as is observed in the lower left quadrant of Fig.\,\ref{fig:plterr}. It follows that our study will focus on nebulae that tend to lie on the high-metallicity end of the data points. As an exercise, we can estimate the range of abundances that the objects in this quadrant cover. To do so, we can  use  the abundance calibration presented in Appendix\,\ref{sec:app}. It was derived from the data used by H06 and consists of a least-squares fit of the observed \oiii\ temperatures as a function of the deduced O/H abundance ratio. When adopting the value of $\tsiii=\toiii=14\,000$\,K as a marker that defines the lower left quadrant, this corresponds to $Z \approx 0.15\,\zsol$ (see Appendix\,\ref{sec:app}). If we select instead the temperature at which the linear regression crosses the dashed line, which is at $\toiii=12\,400$\,K, the inferred metallicity is $Z \approx 0.25\,\zsol$. In summary, the nebulae whose temperatures we attempt to model have metallicities $Z \ga 0.2$\,\zsol.

In standard (steady-state) photoionization models, we disregard hydrodynamical effects and assume that equilibrium ionization and equilibrium temperature are reached at every point in the nebula. In that case, the temperature is determined locally by the condition that the energy is gained  by photoionization and is balanced by the energy loss through line and continuum emission. As one increases the gas metallicity, the equilibrium temperature decreases. A systematic exploration of the \oiii\ and \siii\ temperature behavior with respect to one another was initially carried out as we varied in turn the metallicity $Z$, the ionization parameter ($U$), and the spectral energy distribution (hereafter \sed). It turned out to be a daunting challenge to have \toiii\  significantly exceed  \tsiii. With standard photoionization, we only obtained promising results when we considered combining ionized regions of greatly varying metallicities (Sect.\,\ref{sec:dual}). As an alternative to metallicity inhomogeneities, we subsequently initiated the exploration of three`extraneous' factors: (1) the possible contribution of shock excitation, (2) the presence of temperature inhomogeneities as introduced by Peimbert (1967; c.f. Pe\~na-Guerrero \etal\ 2012a, and references therein), and (3)  the effects of a departure from a Maxwell-Boltzman distribution for the electron energies (Nicholls, Dopita \& Sutherland 2012: hereafter NDS12).

%% METHOD OF SIMPLE STD PHOTOIONIZATION CALC.

\subsection{Photoionization calculations with \map} \label{sec:map}

To compute nebular spectra, we used the multipurpose shock/photoionization code \mapc\ described by Ferruit et\,al. (1997) (see also Binette, Dopita, \& Tuohy 1985), in which the ionization and thermal structure of the nebula are computed in a standard fashion assuming local ionization and thermal equilibria throughout the nebula.  The calculation proceeds at a progression of depths in the nebula until the entire ionization radiation has been absorbed. The updated version of the current code (\mape) now includes the option of considering a non-Maxwell-Boltzmann distribution of the electron energies when calculating excitation rates, as described in Sect.\,\ref{sec:kap}.

\subsubsection{Geometry}\label{sec:geo}
We adopted a simple slab geometry for the ionized gas, since for  the lines of interest the integrated line spectrum does not appreciably differ from that calculated with spherical geometry (Shields 1974). This is the geometry preferred for the Orion nebula that Baldwin et\,al. (1991) modeled in detail. \hii\ regions are very chaotic in appearance, showing evidence of strong density inhomogeneities, with much of the emission coming from condensations or ionized ``skins'' of condensations. Modeling this level of complexity is beyond the current means of photoionization models and is undesirable, unless the observations were sufficiently detailed to constrain such models. Finally, the definition of the ionization parameter in the spherical case depends crucially on the Str\"omgren radius size ($R_s$), which depends on the volume filling factor of the gas and the radius $R_g$ of the empty gas volume near the ionizing stars, both of which are poorly constrained if not downright ill-defined.

We adopted the usual definition of the ionization parameter:
\begin{equation}
U = {\frac{1}{c \nh}} \int_{\nu_o}^\infty {\frac{L_\nu}{4 \pi R_{g}^{2} \, h \nu}}
\,d\nu  = {\frac{\varphi_H}{c \nh}} \; ,  \label{eq:upar}
\end{equation}
\noindent  were $c$ is the speed of light, $h$ the Planck constant, and $R_g$ the distance of the slab from the ionizing star, in other words, \up\ is the ratio between the density of ionizing photons impinging on the slab ($\varphi_H/c$) and the slab H density \nh. With a slab geometry, we assume that the geometrical thickness $R_s-R_g$ of the ionization-bounded slab is insignificant with respect to $R_g$.

In our grid of photoionization calculations, we  varied \up\ over a wide range from 0.001 to 0.046, although in practice a more limited range may suffice (e.g., Evans \& Dopita 1985). Campbell (1988) found that most of the nebulae in her sample lie in the range $-1.9 \ga \lg\up \ga -2.5$. All photoionization calculations were ionization-bounded and we assumed the low-density regime, by adopting $\nh=10\,$\cmc. Some emission lines of low critical density, such as \siiw\ or \oiiw, would be affected by  collisional deexcitation in those few cases where the nebular densities reach values as high as $\sim 500$\,\cmc. Nevertheless, since the overall energy budget remains set by the majority of other lines with much higher critical density,  the conclusions we reached about the \toiii\ and \tsiii\ temperatures do not depend on the density for the overwhelming majority of \hii\ regions.

\subsubsection{Reference set of abundances} \label{sec:abu}

We express the gas metallicity $Z$ with respect to the solar abundance set of Asplund \etal\ (2006), where He/H=0.085 and the other elements  are
%%which we define as $Z=1\,\zsol$. The abundance set with respect to H is the following:
$\emph{}{\rm C,N,O, Ne,Mg, Si, S, Ar, Ca,Fe}\,: {\rm H} = \\
Z \times  (250, 60, 460, 69, 34, 32, 14, 1.5, 2, 28) \times 10^{-6}\,\zsol$,
\noindent where $Z$ is the metallicity in solar units.
%%We will study the effect of varying metallicities by scaling the above abundances by a uniform multiplicative factor equal to $Z$.
We computed nebular models of varying metallicities covering a range from 1\% solar ($Z=0.01\,\zsol$) to more than twice solar ($Z=2.5\,\zsol$).

\subsubsection{Stellar spectral energy distributions} \label{sec:sed}

We explored different stellar {\sed}s for the ionization source of the \hii\ regions. This did not turn out to be a sensitive parameter. In the models reported below, we compare photoionization models with an \sed\ from a zero-age instantaneous stellar burst assuming a Chabrier (2003) initial mass function with \msup=100\msol, and solar metallicity, obtained from the stellar population synthesis models by Bruzual \& Charlot (2003). The stellar atmospheres used to derive the stellar ionizing \sed\ are taken from the Kurucz spectra library (distributed by Lejeune \etal 1997) for stars with \teff $< 50\,000\,$K, and the Rauch (1997) NLTE model atmospheres for the higher temperature stars (Magris \etal\ 2003; hereafter M03). We also explored the alternative case of a single stellar atmosphere of temperature \teff, using the LTE atmosphere models of Hummer \& Mihalas (1970; hereafter HM70).

%% METHOD OF TEMP. FLUCT. t2

\subsection{Models that consider temperature inhomogeneities} \label{sec:fluc}

It has been proposed that the differences between the temperatures from ORLs and CELs could be due to temperature inhomogeneities (e.g. Peimbert 1967). To evaluate how these inhomogeneities  affected the line ratios, we adopted the temperature inhomogeneity characterization described in Binette \& Luridiana (2000: hereafter BL00), who used different temperatures for the CELs in accordance with the energy separation of the atomic levels involved for each line emission calculation.  BL00 did not assess the physical driver of these inhomogeneities. These authors simply assumed their existence and adopted a simple scheme in which the inhomogeneities corresponded to hot-spots as a result of an unknown heating process, distinct from the photoelectric heating.  These authors considered the global effects of temperature inhomogeneities on the emission lines, based on the statistical approximation introduced by Peimbert (1967), which quantifies the brightness increase or decrease of each line in terms of the departure of the temperature from the average temperature \tmean. For each transition, Peimbert used an expansion in Taylor series of the temperature that takes the functional dependence with temperature of each line emissivity into account.
Following Peimbert (1967), the mean nebular temperature \tmean, is defined as
\begin{equation}
\tmean =\frac {\int_V n_e^2 T dV}{\int_V n_e^2 dV} \;  \label{eq:to}
\end{equation}
\noindent  for a homogeneous
metallicity nebula characterized by small temperature inhomogeneities,
$n_e$ is the electron density, $T$ the electron temperature, and
$V$ the volume over which the integration is carried out. The rms
amplitude $t$ of the temperature inhomogeneities is defined as
\begin{equation}
t^2 \equiv \frac {\int_V n_e^2 (T - \tmean)^2 dV}
{\tmean^{2}\int_V n_e^2 dV} \; . \label{eq:tsq}
\end{equation}
\noindent We simplified the expression presented by
Peimbert (1967), in which \tsq\ depends on the density of the ionic species considered ($n_i$) while in the above equations we
implicitly consider only ionized H (by setting $n_i= n_{H^+} = n_e$).

In photoionization calculations, one determines  the \emph{local} equilibrium temperature at every point in the nebula, \teq, that
satisfies the condition that the cooling by radiative processes equals the heating due to the photoelectric effect. Assuming the existence of temperature inhomogeneities and adopting  the scheme of BL00 that describes these as hot-spots, we can define \teq\ as the temperature floor above which putative temperature inhomogeneities take place. These inhomogeneities are taken into account only in the statistical sense, that is, through the use of appropriately `biased' temperatures that depend on the mean squared amplitude of the inhomogeneities and on the functional dependence of the (de)excitation rate on temperature. In this context, \tsq\ also represents the amount of external energy that the hot spots contribute to the nebular energy budget (BL00). Computing the effects of the hot-spots on the line intensities is a two-step process: first  \tmean\ is derived, and  second  the biased temperatures corresponding to each atomic transition are calculated.  For the hot-spot scenario that BL00 simulated, the following expression for \tmean\ provided a satisfying parametric fit
\begin{equation}
\tmean \simeq \teq [1+\gamma(\gamma-1) \,t^2/2]^{-1/\gamma} \; . \label{eq:tmean}
\end{equation}
A value of $\gamma$ of ${\simeq}-15$ was found to provide an adequate parametrization of the behavior of \tmean.

To compute individual recombination line intensities in the presence of small inhomogeneities (which are proportional to
$T^{\alpha}$), the code uses a biased mean temperature \trec\ given by
\begin{equation}
\trec = \langle T^\alpha\rangle^{1/\alpha}\simeq
\tmean \, [1+\alpha(\alpha-1) \,t^2/2]^{1/\alpha} \; . \label{eq:trec}
\end{equation}
Because temperature inhomogeneities have a lesser impact on recombination lines than  on collisionally excited lines, we adopted the simplification of considering a single constant value of the power $\alpha = -0.83$ for all recombination processes.

For collisionally excited lines, when evaluating the excitation rates ($\propto
T^{\beta_{ij}}\, {\rm exp}[-\Delta E_{ij}/kT] $) and deexcitation
($\propto T^{\beta_{ji}}$) of a given multi-level ion, each rate
$ij$ (population) or $ji$ (depopulation) is calculated using
\tmean\ (instead of \teq), and then multiplied by the appropriate
correction factor, either
\begin{eqnarray}
{\rm cf}^{exc.}_{ij} =
1+  \Bigl[ (\beta_{ij} - 1 )
\Bigl(\beta_{ij} + 2 \frac{\Delta E_{ij}}{k_B \tmean} \Bigr) \nonumber
\\ + \Bigl(\frac{\Delta E_{ij}}{k_B \tmean}\Bigr)^2
\Bigr] \frac{t^2}{2}  \label{eq:exc}
\end{eqnarray}
\noindent in the case of excitation, or
\begin{eqnarray}
{\rm cf}^{deexc.}_{ji} =
 1 +  \beta_{ji} (\beta_{ji} - 1 )
\frac{t^2}{2} \; , \label{eq:deexc}
\end{eqnarray}
\noindent in the case of deexcitation. $k_B$ is the standard Boltzmann constant and $\beta_{ji}=\beta_{ij}$. These factors, adapted from the work of Peimbert et~al. (1995),  were applied to all collisionally excited transitions calculated by \mape. It typically results in a significant enhancement of collisional rates by an amount that strongly depends  on the ratio ${\Delta E_{ij}}/{k \tmean}$ for the selected transition $ij$.

%% METHOD OF Z INHOMOGENEITIES

\subsection{Models that consider abundance inhomogeneities} \label{sec:dual}

We explored higher complexity  photoionization calculations in the form of small-scale metallicity inhomogeneities within the ionized gas. One could imagine, for instance, enriched condensations embedded in a more diffuse gas of lower metallicity.  A justification for such a scenario was provided by Tenorio-Tagle (1996). It is based on the fact that the metal-rich ejecta from type\,II supernovae (SNe) ought to follow a long excursion into the galactic environment before they are able to mix with the ISM. In this scenario, the violently ejected matter generates large-scale superbubbles, which eventually cool non-uniformly, creating in the process multiple re-pressurizing shocks. This leads  to the formation of small, dense "cloudlets" of metal-rich gas, which eventually fall back toward the disk of the galaxy because of gravity. An efficient metallicity mixing with the surrounding unenriched ISM will only take place when these ``metallic droplets'' (or cloudlets) are caught within the nebulae associated to star-forming regions.  Stasinska \etal\ (2007) quantitatively investigated the  hypothesis that the photoionization of these droplets can explain the ADFs observed in \hii\ regions. The authors derived bounds of $10^{13}$-–$10^{15}$\,cm on the droplet sizes to (1) explain that they have not yet been spatially resolved and (2) to ensure that they survive long enough their destruction by diffusion.

The possibility of metallicity inhomogeneities  in \hii\ regions has been studied on two scales: (1) on a scale in which they have not yet been spatially resolved\footnote{The alternative of `\emph{large}-scale' metallicity variations has also been studied by Torres-Peimbert, Peimbert \& Pe\~{n}a (1990) and Kingdon \& Ferland (1998). Interestingly, for the case of `\emph{small}-scale' metallicity inhomogeneities, Zhang, Ercolano \& Liu (2007) showed that a small amount of cold gas is sufficient to account for the observed ORL intensities without the need of reproducing the high \tsq\ values `empirically' inferred from \hii\ regions.}  (Tsamis \etal\ 2003), and (2) on a scale associated with proplyds observed in \hii\ regions (Tsamis \etal\ 2011). An example of case (1) is given by the work of  Tsamis \& P\'{e}quignot (2005), who modeled the line spectrum of the central region of 30\,Doradus in the LMC in detail. Their dual metallicity model consists of gas of approximately 'normal' LMC composition (0.36\,\zsol)  co-extensive with a hydrogen-poor component, in which most elements are enhanced by $\approx 0.9$\,dex. The relative weight of the hydrogen-poor component is relatively low, at an 8\% level of the integrated \hb\ flux. In addition to successfully fitting the ORLs from CNO elements (in addition to the UV/optical CELs), the model of Tsamis \& P\'{e}quignot leads to greatly improved predictions for several other optical and IR lines as well. For planetary nebulae (hereafter PNe), the idea has been studied earlier (Liu \etal\ 2000;  Ercolano \etal\ 2003; Tsamis \etal\ 2004; Bohigas 2009; Yuan \etal\ 2011), although the enriched gas component has a totally different origin than that envisaged by Tenorio-Tagle (1996).

To study abundance inhomogeneities, we adopted a dual metallicity modeling approach in which we combined pairs of isobaric photoionization models that simply differ in metallicity. Although we did not consider all physical implications of embedded condensations, such as an intermingled diffuse field, thermal conduction, or a change in ionization parameter, our calculations suffice in evaluating how abundance inhomogeneities can affect the \siii\ and \oiii\ temperatures.  The technique we employed is the following: for a given pair of photoionization models, A and B, we multiplied all line fluxes of model\,A by a weight factor $w_i$  and added them to the corresponding fluxes from model\,B after multiplying these by the complementary weight $1-w_i$. Subsequently, we calculated the line ratios of the resulting `averaged' model and derived physical quantities of interest, such as the 'apparent' \tsiii\ and \toiii\ temperatures.  We built sequences of these models in which only $w_i$ is varying along the sequence. We explored the case where models\,A and B differ only in their metallicity and the case where they differ in both $\up$ and $Z$. Since the results turned out to be qualitatively similar in our temperature diagram (provided the contrast in \up\ was $\la 10$), we will only show cases where both models A and B share the same value in \up. Our sequences can be defined by three  parameters: the ionization parameter \upa, the metallicity $Z_A$ of the (lower abundance) model\,A, and the metallicity contrast between models\,A and B given by $\theta=\lg(Z_B/Z_A)$.

%% METHOD OF k-ELECTRON ENERGY DISTRIBUTION

\subsection{Models with a non-Maxwell-Boltzmann energy distribution}\label{sec:kap}

Recently, NDS12 explored the possibility that electrons in \hii\ regions and PNe depart from a Maxwell-Boltzmann equilibrium energy
distribution. To represent the electron energy distributions, these authors adopted the parametrization given by a $\kappa$--distribution, which is a generalized Lorentzian distribution initially introduced by Vasyliunas (1968). It is widely used in the studies of solar system plasmas (Livadiotis \& McComas 2011; Livadiotis \etal\ 2011) where evidence of suprathermal electron energies abounds. NDS12 found that a $\kappa$--distribution of electron energies resolved many of the difficulties encountered when attempting to reproduce the temperatures observed in nebulae. They concluded that mild departures from a Maxwell-Boltzmann distribution is sufficient for reproducing the spread in temperature values found in diagnostic diagrams such as \trec\ vs. $\left<T_{OII}+T_{NII}\right>$ or $T_{SII}$ vs. $T_{OII}$ in \hii\ regions or \trec\ vs. \toiii\ in PNe.
%For the five temperature diagnostics line ratio diagrams that they analyze, KDS12 were able to account for spread in the data and for the temperature and metallicity discrepancies revealed in their data base.
The $\kappa$--distribution is a function of temperature and $\kappa$. In the limit as $\kappa \rightarrow \infty$, the distribution becomes a Maxwell-Boltzmann distribution. From examining data from \hii\ regions and PNe, NDS12 found that $\kappa \ga 10$ is sufficient to account for most objects they compared their calculations with. These authors proposed various mechanisms that would result in a suprathermal distribution, such as (1) magnetic reconnection followed by the migration of high-energy electrons along field lines, and by the development of inertial Alfv\'en waves, (2) the ejection of high-energy electrons from the photoionization process itself (when the ultraviolet source is very hard, as in PNe or AGN), (3) or from photoionization of dust (Dopita \& Sutherland 2000), and (4) by X-ray ionization, resulting in highly energetic ($\sim 1\,$keV) inner-shell (Auger process) electrons (e.g., Aldrovandi \& Gruenwald 1985).

To implement this new option in \mape, we proceeded as follows. In the case of collisionally excited lines of a given multi-level ion or atom (including \hi\ and \heii), each excitation rate $ij$ is calculated first, using the collision strength $\Omega_{ij}$, the excitation energy $E_{ij}$, and the local equilibrium temperature $T=\teq$, as is customary (Osterbrock 1989). Next, each rate is multiplied by the corresponding correction factor
%\begin{eqnarray}\label{eq:kapa}
\begin{align}\label{eq:kapa}
{\rm cf}^{exc.}_{ij} & =  \frac{\Gamma(\kappa+1)}{\Gamma(\kappa-\frac{1}{2})}\,  \left(1-\frac{3}{2\kappa} \right) \, \left(\kappa-\frac{3}{2}\right)^{-3/2} \\
\nonumber & \times \, \exp\left[\frac{\Delta E_{ij}}{k_B T}\right] \, \left( 1+\frac{\Delta E_{ij}}{(\kappa-\frac{3}{2})\, k_B T} \right) ^{- \kappa}\; ,
\end{align}
%\end{eqnarray}
\noindent which is an expression taken from the work of NDS12 that corresponds to the ratio of the rate computed assuming a $\kappa$--distribution over the rate that would be derived assuming a Maxwell-Boltzmann distribution instead. To prevent numerical overflows, the natural logarithm of Eq.\,\ref{eq:kapa} is evaluated first. To compute the function $\Gamma(z)$, we adopted the expression (Nemes 2010)
\begin{eqnarray}\label{eq:kapc}
\ln \Gamma(z)\eqsim  \frac{1}{2} \left[\ln(2\pi)-\ln(z)\right] + z\, \left[ -1+\ln\left(z+\frac{1}{12z-\frac{1}{10z}}\right)\right]   \ .
\end{eqnarray}
%From an examination of data from \hii\ regions and PNe, it appears that $\kappa \ga 10$ is sufficient to encompass nearly all objects.
To derive the correction factors ${\rm cf}^{deexc.}_{ji}$ for the deexcitation rates $ji$, we used the above expression, but evaluated using $\Delta E_{ji}=0$. This coefficient's value (${\rm cf}_0={\rm cf}^{deexc.}_{ji}$) is unique for a given combination of $\kappa$ and $T$.  It can be applied to threshold-free processes when their cross section  approximately behaves as $\propto E^{-1}$, as is the case with  recombination rates. However, rather than applying correction factors to the recombination rates, we opted to use an effective recombination temperature \trec\ that would result in the same rate coefficients as those obtained by applying the correction factor ${\rm cf}_0$. Since recombination rates approximately vary as $\propto T^{-1/2}$, it follows that the effective temperature to be used when computing recombination rates is given by $\trec \simeq T/({\rm cf}_0)^2$.

We also implemented the enhancement of collisional ionization that can result from $\kappa$--distributions. There are two sets of correction factors: one applies to the excitation-autoionization process (E-A), and the other to direct collisional ionization process (D-I). In the case of E-A, we found from the work of Owocki \& Scudder (1983) that Eq.\,\ref{eq:kapa} can be used to derive the correction factors ${\rm cf}^{auto.}_{ij}$ when the excitation energy is replaced by the relevant autoionizing energies. These coefficients are then applied to the E-A rates. In the case of D-I,  \mape\ uses the method proposed by Arnaud \& Rothenflug (1985) to derive the ionizing rates, which is based on a four-parameter functional fit of the D-I cross-sections $\sigma(E_{ij})$ (their Eq.\,1), for which
we lack an analytical prescription that would allow us to compute the correction factors when considering a $\kappa$--distribution.
%Arnaud \& Rothenflug (1985) provided In the case of a Maxwell-Boltzman integration of these rates is straightforward
%we did not succeed in applying the method\footnote{The ${\rm cf}^{ion.}_{ij}$ are in principle given by dividing Eq.\,B3a by Eq.\,B2a, both taken from Appendix\,B in Owocki \& Scudder (1983).} described by Owocki \& scudder (1983).
We therefore opted for a numerical integration. For each electron shell $j$ of ion $i$ we numerically integrated $\sigma(E_{ij})$ assuming the selected $\kappa$--distribution for the electron energies. We divided the result by the same numerical integration, assuming instead a Maxwell-Boltzmann distribution. The resulting ratio defines the correction factor ${\rm cf}^{dir.io.}_{ij}$, which was then applied to the corresponding D-I rate computed with the  Arnaud \& Rothenflug (1985) expression. For the calculations presented here, the enhancement of collisional ionization caused by a $\kappa$--distribution was found to have no impact on the ionization structure of \hii\ regions.

%% METHOD OF ADDING SHOCK HEATING

\subsection{Photoionization models with shock heating} \label{sec:sho}

To explore the possible impact of shocks in \hii\ regions, we used \mape\ to calculate steady-state plane-parallel shock models, assuming a zero preshock magnetic field. This shock wave, by construction, is set to propagate in a nebula that is \emph{already} photoionized by hot stars. The  nebular gas constitutes the preshock zone with a temperature \tpre\ that is evaluated assuming photoionization equilibrium. We did not include the surrounding unshocked nebular emission in our calculations. Rather than assuming an arbitrary opacity ahead of the shock front, we adopted an unabsorbed ionizing SED. This means that we assumed that the shock wave occurs close to the inner nebula boundary, if homogeneous, or close to the inner boundary of the brightest filaments otherwise. Throughout the calculations, we assumed the low-density limit regime by adopting a density $\npre=1.0$\,\cmc, which we term the preshock density. The basic input parameter of these models will be the postshock temperature \tpost. Using the values of \npre, \tpre\ and \tpost, we computed the Rankine-Hugoniot conditions to determine the shock velocity \vs\ and the postshock density \npost. Both quantities appear in Table\,\ref{tab:sho}. This density \npost\ is about four times that of \npre\ when \vs\ corresponds to a high Mach number.  Downstream of the shock front, as the postshock gas progressively cools, its density increases since the gas pressure is conserved throughout the postshock region. We adopted the previous zero-age instantaneous starburst model of M03 (Sect.\,\ref{sec:sed}) as stellar \sed\ that is responsible for ionizing and heating the preshock zone.  The ionization parameter is defined by Eq.\,\ref{eq:upar} as before, except that \npre\ substitutes \nh.
%Any line emission from the preshock gas is \emph{not} added to the emissivity of the shocked gas reported below.
%The sequences of models  presented in Sect.\,{ref:asho} differ by their metallicity $Z$.

One caveat of our models is, however, that the putative shock wave is set to travel `inward', toward the front face of the nebula, while one would favor the opposite, that is, we expect the shock wave to enter the nebula from the front face (that is, from the side exposed to the incoming stellar UV and stellar wind) and to propagate outward. The reason we are limited to the `inward' direction for the  shock propagation has to do with the way the opacity is integrated, which is subsequently used to determine the radiative transfer of the ionizing flux.
In \mape, this integration starts at the shock front and proceeds downstream in parallel with the temporal evolution of the shocked gas. The opacity is integrated along the flow. This way, the radiative transfer of the radiation emitted within the hot layers  of the shock as well as that from the hot stars can be resolved. However, the opposite `outward' shock propagation might be preferable, for instance in scenarios where shocks are generated by a stellar wind or other stellar ejection mechanisms. In these scenarios where the ionizing photons enter downstream, from the backside of the shock, one  needs to know from the start how the opacity behaves across the entire shocked layer. This requires an iterative method to ensure a proper radiative transfer determination.  Aldrovandi-Viegas \& Contini (1985) developed such an iterative procedure for their code SUMA (Viegas \& Contini 1997) and applied it to model the narrow line region (NLR) of active galactic nuclei (AGN). This option, however, is not available with the code \mape.  To circumvent this limitation, we computed models for which the external ionizing flux is sufficiently intense so that the shocked layer's depth remains a small fraction of the total ionized depth of the nebula in absence of shocks. When this condition is satisfied, the shocked layer can be expected to show a similar structure, whether it is propagating inward or outward, with respect to the direction of the dominant radiation field.

To determine the preionization conditions of the gas  before it enters the shock, we assumed equilibrium ionization and equilibrium temperature. This is a straightforward assumption, since the preshock gas is immersed in a radiation field so intense that it is already highly ionized and at or near equilibrium temperature. The ion and electron temperatures were set to be equal in all shock models. Starting at the shock front, the adiabatic cooling (and recombination) of the plasma is followed only until such time as cooling by radiative processes has sufficiently decreased, so that it becomes equal again to the heating due to photoionization. Hence, the line intensities that we computed strictly apply to the shock-heated layers (that is, the out-of-equilibrium layers beyond which steady-state photoionization eventually settles again, but this zone is not computed).  We recall that the models do not include line emission from the photoionized preshock region either.

To facilitate comparisons between photoionized plasma and steady-state shock waves, we will make use of \emph{column densities}. For  the assumed plane-parallel geometry, the line emission measures are what models integrate to derive line luminosities. However, our interest in referring to integrated column densities is that the internal structure (i.e., the behavior with depth) of the ionization, the temperature, or the cumulative line ratios are the same when comparing homologous models and when the quantities of interest in these models are plotted as a function of column depth. In the familiar photoionization case, models with different densities are homologous, provided the ionization parameter \up\ is the same, at least in the regime of sufficiently low densities where collisional deexcitation does not affect the thermal structure. In the case of shocks, constant \vs\ models (in the low-density regime) are homologous, and their structure remains unchanged as a function of column density, independently of the postshock  density \npost.  The reason is that although the internal energy of the shocked gas after the shock discontinuity scales as \npost, the local cooling rate scales as the density squared, which means that the cooling will proceed on a timescale that evolves as $\nh^{-1}$, that is, decreasing with increasing postshock density. This keeps the behavior of the temperature or ionization invariant as a function of depth, when expressed as a column density. In the case of combined shock-photoionization models, models will be homologous if both \up\ and \vs\ are kept constant when changing the input density.  In the case of pure photoionization, the integrated column density of ionized H for an ionization-bounded model is given\footnote{It is approximately given by $\Nhptot\ \simeq 1.15\,10^{23} \up \,(T/10^4\rm{K})^{0.83}$\cms.} by $\Nhptot \simeq U c/\alpha_B$, where $\alpha_B$ is the case-B recombination rate coefficient of hydrogen.  For the ionization parameter value of \up=0.01 that we assumed in Sect.\,\ref{sec:asho} for the combined shock-photoionization models, $\Nhptot$ is $8.2 \, 10^{20}$\,\cms\ in the pure photoionization case with $Z=1.0\,$\zsol\ (in absence of shocks). In the case of a steady-state shock wave,  \Nhpsho\ increases monotonically with shock velocity\footnote{Dopita \& Sutherland(1996) proposed the following relationship for a near solar metallicity plasma and in absence of magnetic field:
$\lg(\Nhpsho/\cms) \approx 16.852 +5.625 \lg(\vs/100\,\kms) -0.5688 \lg^2(\vs/100\,\kms)$.} \vs\ without external photoionization. It is a manifestation of the longer elapsed time required for the shocked plasma to recombine and cool from \tpost\ to $\sim 100$\,K. In our calculations, for which the shock wave was immersed in a photoionized plasma, $\Nhpsho$ when integrated over the out-of-equilibrium layers only, is as low as $2.5\, 10^{19}$\,\cms\ for a shock velocity of 56\,\kms. Hence, the aforementioned condition of considering only shocked layers that are geometrically thin with respect to an ionization-bounded photoionized layer is clearly satisfied with the velocities considered in this work ($\le 65\,$\kms).

\begin{figure}[ht]
\begin{center} \leavevmode
\includegraphics[width=0.49\textwidth]{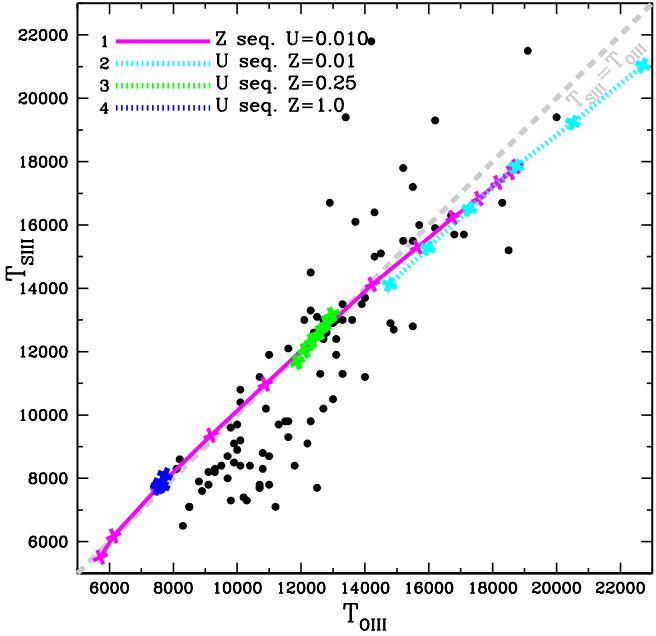}
\caption{Photoionization calculations with an \sed\ corresponding to a zero-age instantaneous starburst model from M03 with $\msup=100\,$\msol. A sequence of models of increasing metallicity $Z$ is represented by the  magenta solid line, for which $\up$ is constant at 0.01. The  metallicities covered a range from $Z=0.01\,\zsol$ to $Z=2.5\,\zsol$, where $Z$ increases by 0.2\,dex from model to model.  Three sequences of models are superimposed in which \up\ is increased (rather than $Z$), covering the range 0.001 to 0.046. Each sequence has a different metallicity, as follows: $Z=0.01\,\zsol$ (cyan dotted line), $Z=0.25\,\zsol$ (light green dotted line) and $Z=1.0\,\zsol$ (blue dotted line). In this and subsequent figures, a light gray dashed line represents the locus of equal temperatures.}
\label{fig:figh1}
\end{center}
\end{figure}

%%%%%%%%%%%%%%%%%%%%%%%%%%%%%%%%%%%%%%  RESULTS  %%%%%%%%%%%%%%%%%%%%%%%%%%%%%%%%%

\section{Model results and discussion}\label{sec:res}

In figures without the error bars, such as Fig.\,\ref{fig:figh1}, the span in \oiii\ temperatures to the right of the dashed line in the lower left quadrant stands out more clearly. At the low-temperature end, \toiii\ can exceed \tsiii\ by as much as $\sim 3\,000$\,K.  We now investigate which types of models or parameter values would cause the \oiii\ temperatures to be appreciably higher than those from \siii. We begin with homogeneous photoionization calculations.

\newpage
%% RESULTS FROM STD SIMPLE PHOTOIONIZATION

\subsection{The pure homogeneous photoionization case}\label{sec:pure}

When the plasma temperature is set by the condition that equilibrium ionization and equilibrium temperature apply locally everywhere within the nebula, and that there are no inhomogeneities in either the local densities or the metallicities, one finds that varying the input parameters gives little leverage on the behavior of both temperatures. This is illustrated by the models presented in Fig.\,\ref{fig:figh1}. The magenta solid line shows a sequence of 13 models (all with $\up=0.01$ and the zero-age instantaneous starburst \sed\ of M03), in which only the metallicity increases along the sequence, by 0.2\,dex from model to model, covering the range $Z=0.01$ to 2.5\,\zsol. Higher metallicity models result in cooler nebulae, while the lowest $Z$ models occupy the upper right corner. What we infer from this plot is that \emph{below} 14\,000\,K, there is no significant departure of the models from the equal temperature (dashed) line. Only beyond 16\,000\,K does a modest departure from the equal temperature line begin to occur. To complete the picture, we show  three sequences, of six models each,  starting at $\up=0.001$ and ending at $0.046$, along which the ionization parameter increases by 0.33\,dex from model to model. These three sequences are characterized by different metallicities $Z$ of 0.01\,\zsol\ (cyan dotted line), 0.25\,\zsol\ (light green dotted line) and 1.0\,\zsol\ (blue dotted line), respectively. They illustrate quantitatively that \up\ is not a significant parameter, except at very low metallicities. In this case, the value of \up\ matters because of the cooling provided by collisional excitation of \hi, which plays an increasingly important role in establishing the equilibrium temperature when metal cooling becomes progressively inoperative beyond $Z\la 0.05$\,\zsol. Increasing \up\ reduces the neutral fraction of hydrogen H$^0$/H, which diminishes the cooling that \hi\ collisional excitation can provide, resulting in a hotter nebula.
%Even then, the departure from the dashed line is modest.

\begin{figure}
\begin{center}
\leavevmode
\includegraphics[width=0.49\textwidth]{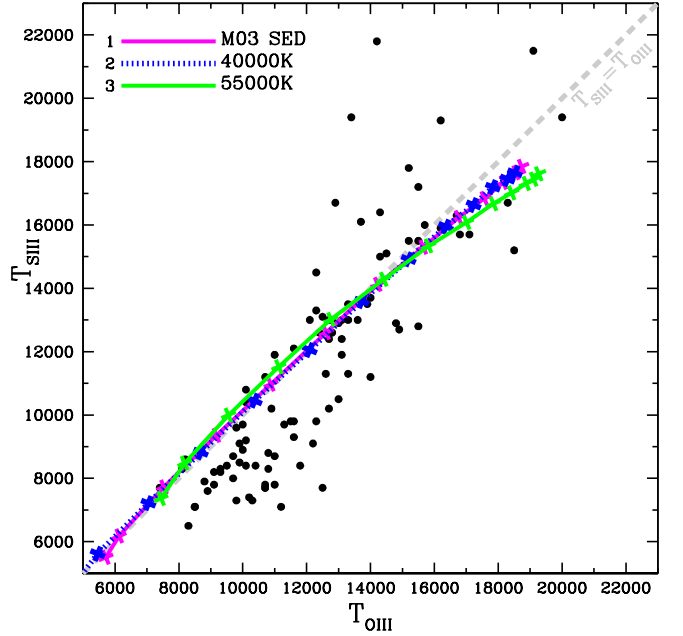}
\caption{
Comparison of two sequences in metallicity that differ by the shape of their ionizing \sed: a single temperature star of either $\teff=55\,000$\,K (light green line) or $\teff=40\,000$\,K (blue dotted line).   The previous zero-age instantaneous starburst \sed\ of M03 with $\msup=100\,$\msol\ appears underneath (magenta line). Each sequence comprises 13 models with the metallicity increasing by a factor 0.2\,dex from model to model, starting at $Z=0.01\,\zsol$ and ending at 2.5\,\zsol. The ionization parameter is $\up=0.01$ in all calculations.  }
\label{fig:figh2a}
\end{center}
\end{figure}

In Fig.\,\ref{fig:figh2a}, we compare the case of using an \sed\ consisting of a single star atmosphere model (HM70) with $\teff=55\,000$\,K (light green solid line) with the case of using the previous M03 \sed\ (magenta line). Both sequences are metallicity sequences covering the interval $Z=0.01$ to 2.5\,\zsol. We also explored stellar {\sed}s of lower \teff\ or  {\sed}s from M03 of lower \msup\ and/or of an evolved population. To prevent cluttering in the figure with too many overlapping lines, these models are not plotted, except for the $\teff=40\,000$\,K HM70 stellar atmosphere \sed\ (blue dotted line). Those models showed that reducing either \msup\ or \teff\ does not cause any improvement in reproducing the spread in the observed \tsiii\ and \toiii\ temperatures of the lower left quadrant.
The {\sed}s that we adopted for the calculations shown in Fig.\,\ref{fig:figh2a} illustrate that a change in the continuum hardness is not effective in resolving the \tsiii--\toiii\ temperature issue.

%%% RESULTS FROM TEMPERATURE INHOMOGENEITIES

\subsection{Results of models with temperature inhomogeneities}\label{sec:atsq}

\begin{figure}[ht]
\begin{center} \leavevmode
\includegraphics[width=0.49\textwidth]{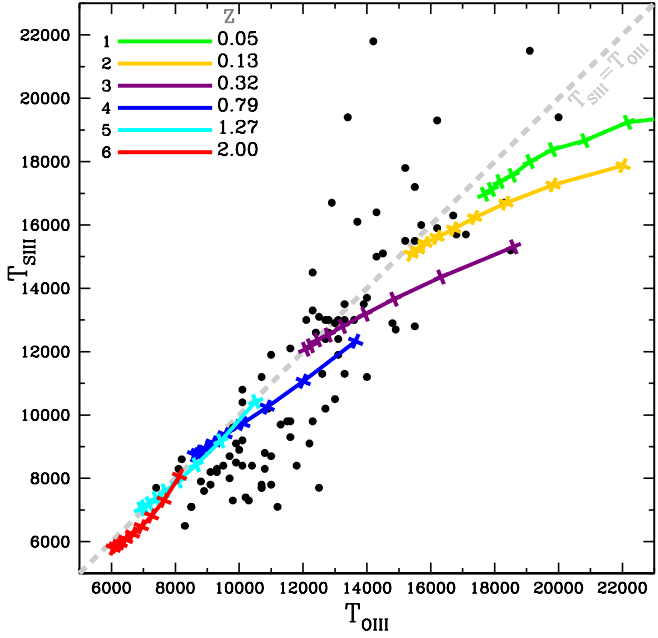}
\caption{Sequences of models in which \tsq\ is increased from 0.004 to 0.16 by successive factors of 1.58 ($=0.2$\,dex). All calculations were carried out with $\up\ = 0.01$ and the zero-age instantaneous starburst model \sed\ with $\msup=100\,\msol$ from M03. Each sequence corresponds to a different metallicity $Z$, which is indicated in the line color legend in \zsol\ units.
}
\label{fig:figh3aa}
\end{center}
\end{figure}

Because \mape\ offers the possibility of simulating the effects of temperature inhomogeneities following the method described in Sect.\,\ref{sec:fluc}, we  explored various runs of models in which the mean squared amplitude of these inhomogeneities (\tsq) is progressively increased\footnote{Note that \tsq\ as implemented in Sect.\,\ref{sec:fluc} is defined via a second-order expansion of a Taylor series and as such is accurate only for low values of $\tsq < 0.1$.}. The \tsq\ values that we explored range from 0.004 to 0.16, fully encompassing that found in \hii\ regions. For instance, in the sample of 28 nebulae used by Pe\~na-Guerrero, Peimbert \& Peimbert (2012b, hereafter PG12) to recalibrate the Pagel method, 70\% of the objects have \tsq\ values within the interval $0.02 \leq \tsq\ \leq 0.04$, one object has a lower value, while the rest (25\%) lie within the range 0.06--0.12. The results of our calculations are shown in Fig.\,\ref{fig:figh3aa}. All calculation were carried out with \up\ = 0.01 and the starburst \sed\ from M03. Each \tsq--sequence comprises nine models, with \tsq\ increasing by a factor 0.2\,dex (=1.58) from model to model. The different \tsq-sequences differ only in their metallicity, whose value appears in \zsol\ units in the legend of Fig.\,\ref{fig:figh3aa}. We found that the modifications to the procedure for calculating emissivities (with respect to pure photoionization) suffice to explain the growing divergence of low-metallicity sequences from the equal temperature line as \tsq\ is increased. Changes in ionization structure due to calculating the recombination rates at \trec\ (see Eq.\,\ref{eq:trec}), rather than at the equilibrium temperature, are minor.

Figure\,\ref{fig:figh3aa} shows that, for models within the lower left quadrant, increasing \tsq\ increases the temperatures \tsiii\ and \toiii\ by comparable amounts. As a result, the \tsq--sequences do not cover the spread in the data observed to the right of the equal temperature line in the lower left quadrant. Only within the upper right quadrant do the calculated temperatures diverge from the dashed line as \tsq\ is increased. Hence,  temperature inhomogeneities cannot account by themselves for the excess in \toiii\ temperatures observed in the lower left quadrant.

%% RESULTS FROM METALLICITY INHOMOGENEITIES

\subsection{Results of models with metallicity inhomogeneities}\label{sec:met}

\begin{figure}
\begin{center} \leavevmode
\includegraphics[width=0.49\textwidth]{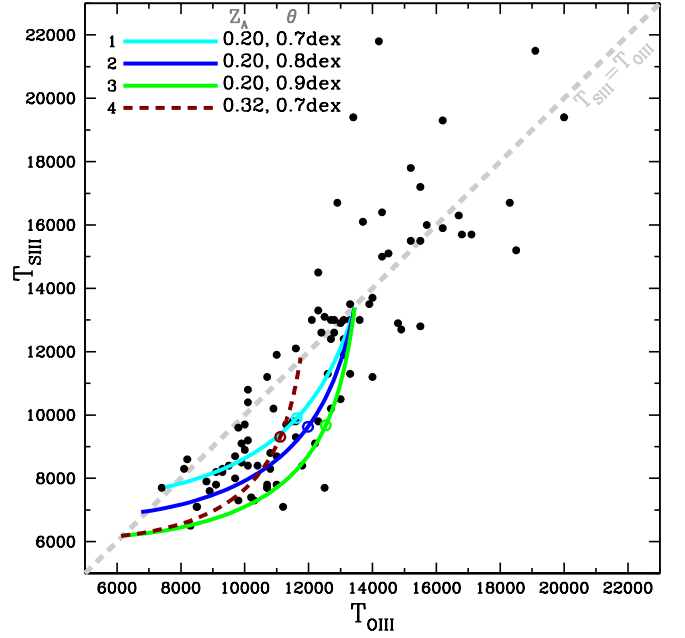}
\caption{Sequences of models in which the relative weight $w_i$ of two photoionization calculations is varied, as described in Sect.\,\ref{sec:dual}. For each sequence, the metallicity of its model\,A in solar units appears in the line color legend followed by the contrast $\theta$ in dex. An open circle indicates the position where the relative weight applied to model\,A and B is equal ($w_i=0.5$). In all the above models, the ionization parameter is $\up=0.01$ and the \sed\ corresponds to the zero-age instantaneous starburst model with $\msup=100\,\msol$. }
\label{fig:met}
\end{center}
\end{figure}

%more complex photoionization models that consider abundance inhomogeneities,
We now turn to the dual metallicity modeling approach described in Sect.\,\ref{sec:dual}, which has the advantage of not requiring an external contribution to the energy balance of the nebula. We report on sequences of models that have identical values in ionization parameter $\upa=\upb=0.01$. Reducing this parameter to 0.001 did not affect the results in a noticeable way, except in cases where $Z_A$ was $\ll 0.2\,\zsol$. We recall that along any sequence, only the relative weight $w_i$ applied to models\,A and B varies, within the interval $0 \leq w_i \leq 1$. In Fig.\,\ref{fig:met} we present examples of dual metallicity sequences. The $\theta=0.7$\,dex sequence (cyan line) corresponds to a weighted average of a model A with $Z_A=0.2\,\zsol$ with a model\,B with $Z_B=1.0\,\zsol$. The blue line sequence and the light green line sequence have the same $Z_A$, but a higher abundance contrast $\theta$ of 0.8 and 0.9\,dex, respectively.  The maroon dashed line sequence was calculated with a higher value $Z_A=0.32\,\zsol$ and a metallicity contrast of 0.7\,dex. We infer from Fig.\,\ref{fig:met} that  sequences with a sufficiently high contrast $\theta$ can cover the observed domain of points to the right of the equal temperature (dashed) line. The locus of the various models also suggests that a metallicity contrast extending up to $\la 1$\,dex is required to account for the nebulae with the largest temperature differences of $\toiii-\tsiii$. A possible drawback might be that the modeling relies on some fine-tuning of $w_i$ to reach the observed \toiii\ value. To give an idea of how \tsiii\ and \toiii\ varies with $w_i$, we use an open circle to indicate the position along each sequence where both models A and B have equal weight values of 0.5. This corresponds to the case where models\,A and B reprocess the same amount of ionizing photons (and emit about the same flux in recombination lines).

\begin{figure}
\begin{center} \leavevmode
\includegraphics[width=0.49\textwidth]{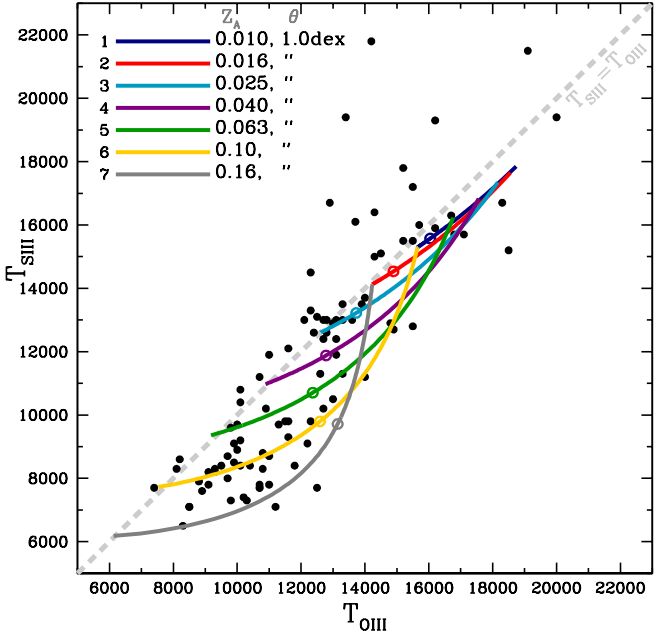}
\caption{Seven sequences of dual metallicity models covering a wider range in metallicities. The  metallicity contrast between $Z_B$ and $Z_A$ is fixed at $\theta=1.0$\,dex. Each sequence corresponds to a different metallicity $Z_A$ value, which is indicated in the line color legend in \zsol\ units.  An open circle along each sequence indicates the position where the relative weight applied to each model is equal ($w_i=0.5$). In all the above sequences, the ionization parameter is $\up=0.01$, and the \sed\ corresponds to the zero-age instantaneous starburst model with $\msup=100\,\msol$. }
\label{fig:lmet}
\end{center}
\end{figure}

One interesting feature of models that consider abundance inhomogeneities is that the divergence from the equal temperature line becomes markedly smaller toward the high-temperature end. The reason is that beyond $Z \la 0.1$\,\zsol, metals play a progressively vanishing role in affecting the plasma temperature. This behavior is illustrated by  the seven sequences that are presented in Fig.\,\ref{fig:lmet}, which extend over a much wider metallicity range compared to the previous figure. The abundance contrast $\theta$ is fixed at 1.0\,dex and the metallicity of the model\,A of each sequence appears in \zsol\ units in the figure legend. Data of higher S/N would be essential to ascertain how nebulae behave in the low-metallicity (high-temperature) quadrant of the diagram.

%%% RESULTS FROM k_DISTRIBUTIONS

\subsection{Results of models with a $\kappa$--distribution}\label{sec:kapm}

\begin{figure}[ht]
\begin{center} \leavevmode
\includegraphics[width=0.49\textwidth]{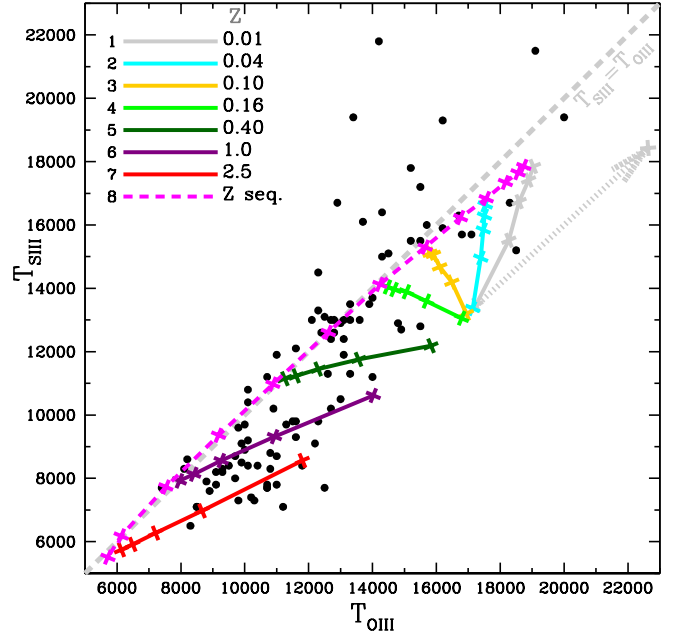}
\caption{Sequences of models in which the parameter $\kappa$ has the values of 80, 40, 20, 10, and 5. The larger $\kappa$, the closer the model to standard models with a Maxwell-Boltzmann electron energy distribution.  These $\kappa$ sequences differ in their metallicity $Z$ (expressed in \zsol\ units in the line color legend). The dashed magenta line represents a \emph{metallicity} sequence of standard photoionization models ($\kappa=\infty$) (same sequence as in Fig.\,\ref{fig:figh1}). The light gray arrow illustrates the increase in the \tsiii\ and \toiii\ temperatures  when \emph{enhanced} \hi\ excitation is left out from the calculations of the model with $Z=0.01$\,\zsol\ and $\kappa=5$. In all calculations, we adopted $\up=0.01$ and an \sed\ consisting of a zero-age instantaneous starburst model from M03 with $\msup=100\,$\msol.}
\label{fig:figh7}
\end{center}
\end{figure}

After implementing the effects of a $\kappa$--distribution on the line intensities and recombination processes as well as on the equilibrium temperature of a photoionized plasma, we proceeded to compute sequences of models where $\kappa$ has the values of 80, 40, 20, 10, and 5 (which encompasses the range 6--20 that NDS12 explored).
In Fig.\,\ref{fig:figh7}, we present seven such sequences that differ only in their gas metallicity. The gas metallicity of each sequence appears in the figure legend, expressed in \zsol\ units. The dashed magenta line represents a metallicity sequence with a pure Maxwell-Boltzmann distribution (the same $Z$-sequence that was plotted in Fig.\,\ref{fig:figh1}). All calculations assumed $\up=0.01$. Our results indicate that $\kappa$--distributions are able to reproduce the spread in the observed values of \toiii\ vs. \tsiii\ inside the lower left quadrant and therefore constitute an interesting alternative to the other options studied in this paper.

With $\kappa$--distributions, the higher the excitation energy of a level, the larger  the increase in the corresponding emission  line intensity. This explains why \oiiitw\ grows considerably more than \siiitw. At a fixed $\kappa$ value, the ratio between the respective \tsiii\ and \toiii\ temperature increases is constant. Hence, independent of metallicities, models with the same $\kappa$ lie at about the same distance from the equal temperature dashed line in  Fig.\,\ref{fig:figh7}. This is also the case for the other temperature diagrams presented by NDS12. Because \mape\ includes the enhancement in \hi\ excitation cooling due to the high-energy tail of $\kappa$--distributions,
\toiii\ and \tsiii\ are seen to increase less and less in Fig.\,\ref{fig:figh7} when decreasing metallicity at a fixed $\kappa$ value. A measure of this effect is represented by the arrow in Fig.\,\ref{fig:figh7}, which represents the increase in the \tsiii\ and \toiii\ temperatures  when \emph{enhanced} \hi\ excitation is left out from the calculations of the model with $Z=0.01$\,\zsol\ and $\kappa=5$. Since collisional excitation of \hi\ is a strong function of the neutral fraction of hydrogen H$^0$/H (Luridiana \etal\ 2003), the importance of the effect described above at low metallicities is governed by the ionization parameter\footnote{The sensitivity of the equilibrium temperature on \up\ was briefly analyzed in Sect.\,\ref{sec:pure} in reference to the \up\ sequence with $Z=0.01\,$\zsol\ (cyan dotted line in Fig.\,\ref{fig:figh1}). With $\kappa$--distributions, this sensitivity is amplified.}. If we were to increase \up, it would result in a lower H$^0$/H fraction and this would reduce the importance of \hi\ excitation cooling present at very low metallicities and therefore result in a higher nebular temperature.

%The magnitude of the temperature increase will depend on \up\ at very low metallicities. This process leads to an increase in the total cooling rate, which results in a lower equilibrium temperature than if \hi\ excitation was not present.

\begin{figure}[ht]
\begin{center} \leavevmode
\includegraphics[width=0.49\textwidth]{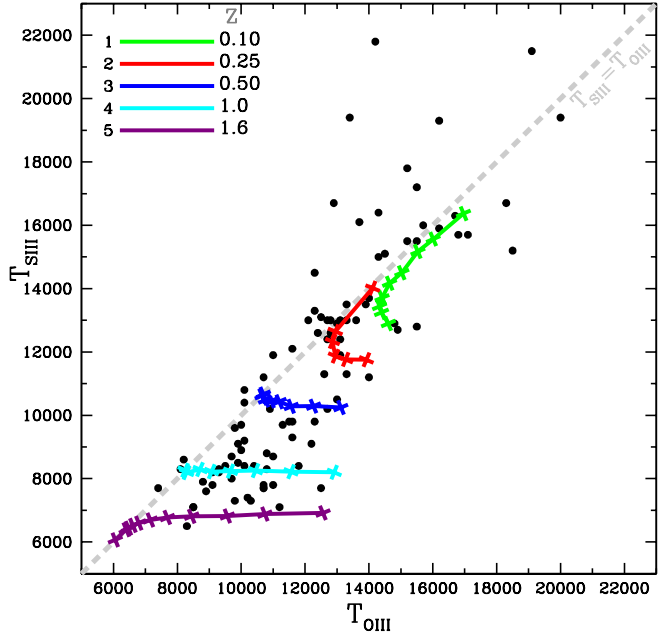}
\caption{Six sequences of combined shock-photoionization models along which the shock velocity \vs\ is increased in locked steps. The preshock conditions are set by photoionization equilibrium, using the parameter $\up = 0.01$ and the M03 \sed, and assuming thermal and ionization equilibria (see Sect.\,\ref{sec:sho}). Each sequence is characterized by a different metallicity $Z$ value, which is expressed in \zsol\ units in the line color legend. For each model, the preshock temperature, the postshock density, the postshock temperature, and the resulting shock velocity are given in Table\,\ref{tab:sho}. The first two models of the $Z=0.5$\,\zsol\ and 1.0\,\zsol\ sequences overlap too closely to be distinguished.}
\label{fig:sho}
\end{center}
\end{figure}

%%% RESULTS FROM SHOCK-PHOTOIONIZATION MODELS

\subsection{Results from photoionization models with shock heating}\label{sec:asho}

A set of calculations was carried out in which we computed the line intensities of a plane-parallel shock wave that propagates within a photoionized gas layer. To simplify the exercise, we adopted  the zero-age instantaneous starburst \sed\ and  the ionization parameter value of $\up=0.01$ for the preshock photoionized gas in all calculations. More information about the models is presented in Sect.\,\ref{sec:sho}.

In Fig.\,\ref{fig:sho} we present  six sequences of combined shock-photoionization calculations, which differ by their metallicity $Z$ (see the adopted values expressed in \zsol\ units in the figure legend). Preshock temperatures (\tpre) are given for each model in Table\,\ref{tab:sho}. Along each sequence, it is the post-shock temperature (hence, the shock velocity) that is progressively increased from model to model.  The exact values of the postshock temperature, \tpost, and the associated shock velocities, \vs,  are listed for each model in Table\,\ref{tab:sho} (along with \npost). The associated shock velocities typically cover the interval 20 to 60\,\kms. This velocity range is very similar to what is obtained in detailed kinematical studies of GEHRs and circumnuclear star-forming regions (see e.g. Rozas \etal\ 2006a, b; Firpo et\,al. 2011; H\"agele et\,al. 2010). The results depicted in Fig.\,\ref{fig:sho} turn out to be quite encouraging.  In the lower left quadrant, the models are surprisingly successful in covering the spread in object positions to the right of the equal temperature (dashed) line. As the value of the sequences'metallicity decreases, the calculated temperatures move closer to the dashed line. A discussion on the behavior of \toiii\ in combined shock-photoionization models can be found in Appendix\,\ref{app:sho}.

%%%%%%%%%%%%%%%%%%%%%%%%%%%%%  TABLE 1   %%%%%%%%%%%%%%%%%%%%%%%%%%%%%%%%%%%%%%%%%%%%
\begin{table}
 \caption{Shock heated photoionization models}
 \label{tab:sho}
 \centering
 \begin{tabular}{cllllll}
 %\begin{tabular}{crcllll}
 \hline\hline
 %\vspace{1mm}
    Sequence & color\tablefootmark{a} & $Z$ & \tpre\ & \npost\ & \tpost\ & \vs\ \\
     &  & ($Z/\zsol$) & (K) & (\cmc) & (K) & \kms \\
    \hline
 %   0    & yellow & 0.01 &  18\,070 & 1.06 & 19\,880 & 20.9\\
 %  \multicolumn{4}{l}{} & 1.28 & 23\,850 & 25.2 \\
 %  \multicolumn{4}{l}{} & 1.62 & 28\,620 & 30.3 \\
 %  \multicolumn{4}{l}{} & 1.96 & 34\,350 & 35.9  \\
 %  \multicolumn{4}{l}{ } & 2.27 & 41\,220 & 41.9 \\
 %  \multicolumn{4}{l}{ } & 2.54 & 49\,460 & 48.2 \\
 %  \multicolumn{4}{l}{ } & 2.77 & 59\,350 & 54.9 \\
    1    & green  & 0.10 & 15\,400 & 1.12 & 16\,940 & 20.2\\
   \multicolumn{4}{l}{ } & 1.45 & 20\,330 & 24.4 \\
   \multicolumn{4}{l}{ } & 1.79 & 24\,390 & 29.1 \\
   \multicolumn{4}{l}{ } & 2.11 & 29\,970 & 34.2 \\
   \multicolumn{4}{l}{ } & 2.40 & 35\,130 & 39.6 \\
   \multicolumn{4}{l}{ } & 2.66 & 42\,150 & 45.4 \\
   \multicolumn{4}{l}{ } & 2.87 & 50\,580 & 51.5 \\
   \multicolumn{4}{l}{ } & 3.06 & 60\,700 & 58.0 \\
   \multicolumn{4}{l}{ } & 3.21 & 72\,840 & 65.0 \\
   2    & red & 0.25  &  12\,500  & 1.19 & 13\,740 &  18.6\\
   \multicolumn{4}{l}{ } & 2.18 & 23\,740 & 31.2 \\
   \multicolumn{4}{l}{ } & 2.71 & 34\,180 & 41.2 \\
   \multicolumn{4}{l}{ } & 3.10 & 49\,230 & 52.5 \\
   \multicolumn{4}{l}{ } & 3.24 & 59\,070 & 58.8 \\
   \multicolumn{4}{l}{ } & 3.37 & 70\,880 & 65.5 \\
   3    & blue & 0.51  & 10\,800  & 1.73 & 14\,260 &  20.1\\
   \multicolumn{4}{l}{ } & 2.06 & 17\,110 & 25.8 \\
   \multicolumn{4}{l}{ } & 2.61 & 24\,630 & 34.4 \\
   \multicolumn{4}{l}{ } & 2.84 & 29\,560 & 39.1 \\
   \multicolumn{4}{l}{ } & 3.03 & 35\,470 & 44.0 \\
   \multicolumn{4}{l}{ } & 3.19 & 42\,570 & 49.4 \\
   \multicolumn{4}{l}{ } & 3.32 & 52\,080 & 55.2 \\
   \multicolumn{4}{l}{ } & 3.43 & 61\,300 & 61.4 \\
   4    & cyan & 1.00 & 7\,400 & 1.33 &8\,140 &  14.8\\
   \multicolumn{4}{l}{ } & 2.31 & 14\,070 & 24.5 \\
   \multicolumn{4}{l}{ } & 2.80 & 20\,250 & 30.8 \\
   \multicolumn{4}{l}{ } & 3.00& 24\,300 & 36.2 \\
   \multicolumn{4}{l}{ } & 3.16 & 29\,160 & 40.6 \\
   \multicolumn{4}{l}{ } & 3.30 & 35\,000 & 45.3 \\
   \multicolumn{4}{l}{ } & 3.42 & 42\,000 & 50.5 \\
   \multicolumn{4}{l}{ } & 3.51 & 50\,400 & 56.1 \\
   5    & purple & 1.58 & 5\,500 & 1.55 & 6\,050 & 13.6\\
   \multicolumn{4}{l}{ } & 2.21 & 8\,712 & 18.9 \\
   \multicolumn{4}{l}{ } & 2.73 & 12\,540 & 24.8 \\
   \multicolumn{4}{l}{ } & 2.93 & 15\,050 & 28.1 \\
   \multicolumn{4}{l}{ } & 3.11 & 18\,060 & 31.6 \\
   \multicolumn{4}{l}{ } & 3.25 & 21\,680 & 35.4 \\
   \multicolumn{4}{l}{ } & 3.38 & 26\,010 & 39.4 \\
   \multicolumn{4}{l}{ } & 3.48 & 31\,220 & 43.8 \\
   \multicolumn{4}{l}{ } & 3.57 & 37\,460 & 48.6 \\
   \hline
\end{tabular}
\tablefoot{
In all calculations reported in the table, we used the zero-age instantaneous starburst \sed\ from M03 with $\msup=100\,$\msol\ and adopted $\up=0.01$ and $\nh=\npre=1.0\,$\cmc\ for the preshock gas.\\
\tablefoottext{a}{The color coding is the same as in Fig.\,\ref{fig:sho}.}
}
\end{table}

Despite the relative success of our calculations, it should be noted that they represent an oversimplification of a very complex problem.  A subsequent analysis with 3D hydrodynamical simulations is necessary. As stated earlier, the line intensities and the temperatures calculated in the models shown in Fig.\,\ref{fig:sho} correspond to the zone that is shock-heated. That is, they include line emission from the shock discontinuity down to the point where cooling and heating (by the UV flux) become equal again, at which point the calculations stop\footnote{Interestingly, there is negligible \oiiiw\ emission taking place beyond that point because of the high density characterizing that zone, resulting in the emission of \emph{low}-excitation emission lines instead.}. As indicated in Sect.\,\ref{sec:sho}, the total column density of an isobaric ionization-bounded photoionization model with $\up = 0.01$ is 32 times that of the shocked layer of the 56\,\kms\ model from the $Z=1.0\,\zsol$ sequence (last model of sequence no.\,4 in Table\,\,\ref{tab:sho}). This is a concern, since it appears to suggest that one might need 32 successive shocks propagating across the whole nebula for the resulting emission lines to reflect the presence of shocks at the level shown in Fig.\,\ref{fig:sho}, a rather implausible proposition. Instead of moving inward, we assumed a possibly more realistic scenario in which the shocks are moving outward as a result of a stellar wind generated by the ionizing stars. In this case, the following scenario emerges. A high Mach number shock results in a compression factor of $\sim 4$ of the pressure (across the shock front) and, from that point on,  the density increases because the pressure is conserved across the adiabatic flow, as the gas cools and the temperature decreases. At the point downstream where photoheating balances cooling again, the density has reached a value of about $4 \tpost/\tpre$, where \tpre\ is an estimate of the local equilibrium temperature.   This amounts to a compression factor of $\sim 28$. Hence the effective ionization factor near the back-photoionized zone is $\sim 3.5\,10^{-4}$, and even if this layer was kept photoionized, its \oiiiw\ emission contribution could not be significant. As the shock wave progresses outward within the nebula, the accumulating  denser gas shell behind the shock will eventually quench the photon flux  responsible for photoionizing the preshock zone ahead. The latter would quickly evolve into a low-excitation \emph{fossil} nebula (ahead of the shock) without significant emission in \oiiiw. If we consider the back of the outward-moving shock (facing the incoming ionizing radiation), we may expect the accumulating dense shell to break up into filaments as a result of instabilities.
%generated by the small net cooling available when photoionization is close to equilibrium conditions.

Another complication that our simple plane-parallel calculations does not consider is that \hii\ regions are characterized by a low volume filling factor (Osterbrock 1989). The fine spatial structure of \hii\ regions has so far not been fully resolved. Although progress has been significant, we have not yet reached the microstructure scale of nebular gas (e.g. O'Dell et\,al. 2003; Mesa-Delgado et\,al. 2008; Mesa-Delgado \& Esteban 2010). It has been proposed that nebulae consist of gas condensations or filaments close to being ionization-bounded (e.g. Tsamis \& P\'equignot 2005). The shock scenario described above can in principle be transposed to the case of dense condensations. 3D hydrodynamical models, such as the one developed by Raga \etal\ (2008), would be required to simulate such a case and to determine the temperature of the diffuse gas in which the large-scale shock surrounding the condensations (or the filaments) propagates.

To summarize, despite the caveats just discussed, we consider that shock-photoionization models have the potential of reproducing the observed spread in temperatures observed below the equal temperature line. These models show a tendency quite similar to that of abundance inhomogeneity sequences presented in Sect.\,\ref{sec:met}, since \toiii\ approaches \tsiii\ as the gas metallicities decrease (toward the upper right quadrant). This behavior is the opposite of that observed with temperature inhomogeneities as modeled with \tsq\ parametrization (Sect.\,\ref{sec:atsq}).

\begin{figure}
\begin{center} \leavevmode
\includegraphics[width=0.49\textwidth]{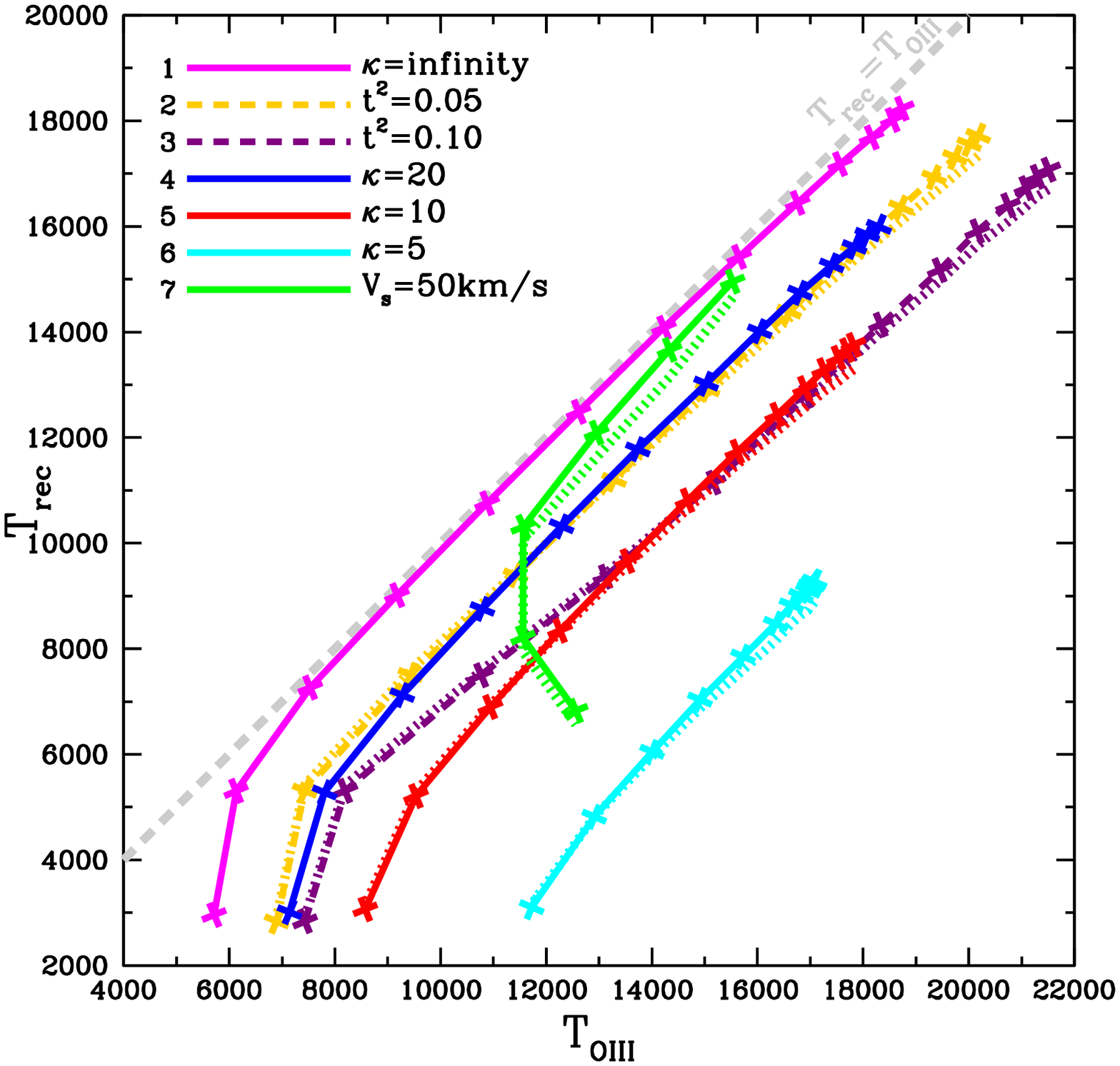}
\caption{Recombination temperature \treco\ as a function of \toiii\ for various models. Along each sequence, only the metallicity varies. Solid and dashed lines are used to represent \treco. A dotted line of the same color depicts \trech, but it is plotted only for cases where \trech\ differs from \treco\ by more than 3\%. The ionization parameter is always $\up=0.01$. In the case of the  50\,\kms\ combined photoionization-shock model sequence (light green), $Z$ assumes the values of 0.01\,\zsol, 0.10\,\zsol, 0.25\,\zsol\, 0.49\,\zsol\, 1.0\,\zsol, and 1.6\,\zsol\ (the higher the metallicity, lower is \trec). All other sequences cover the same metallicity range of 0.01\,\zsol\ to 2.5\,\zsol, with a fixed increment of 0.2\,dex between successive models.  Three sequences represent $\kappa$--distributions (20, 10, or 5, see legend) and two sequences represent the case of temperature inhomogeneities ($\tsq=0.05$ or 0.10, see legend). The homogeneous pure photoionization sequence is represented by the magenta line (same as in Fig.\,\ref{fig:figh1}). }
\label{fig:treca}
\end{center}
\end{figure}

\begin{figure}
\begin{center} \leavevmode
\includegraphics[width=0.49\textwidth]{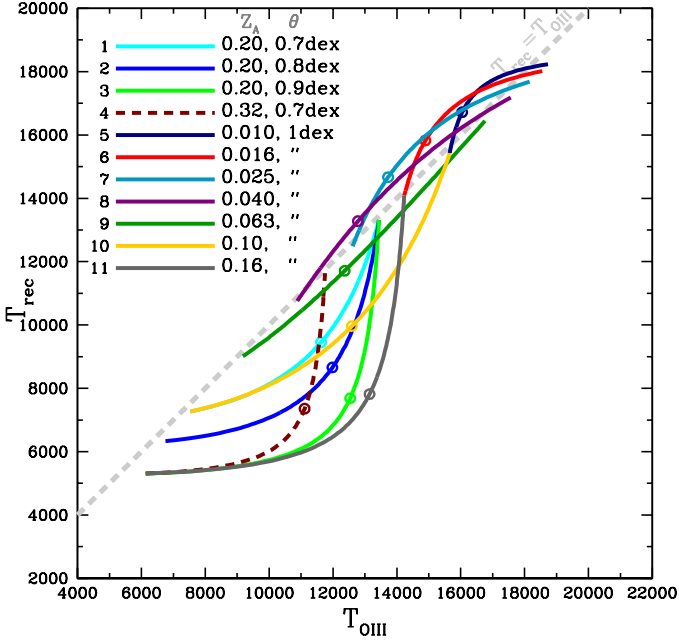}
\caption{Recombination temperatures \treco\ as a function of \toiii\ for models that consider metallicity inhomogeneities as discussed in Sect.\,\ref{sec:met}. Along each sequence, only the metallicity varies (see legend for values of $Z_A$ and $\theta$). In no model does  \trech\ differs from \treco\ by more than 3\%. All four sequences from Fig.\,\ref{fig:met} with varying contrast $\theta$ are included as well as the seven sequences from Fig.\,\ref{fig:lmet} with $\theta=1.0$ (color-coding and nomenclature remain the same, see legend).
%These correspond to a model\,A that has $Z_A=0.01$\,\zsol\ (navy line), 0.04\,\zsol\ (purple), 0.063\,\zsol\ (dark green), 0.10\,\zsol\ (yellow) and  0.16\,\zsol\  (dark gray line), respectively.
An open circle indicates the position where $w_i=0.5$. The ionization parameter is $\up=0.01$ throughout. }
\label{fig:trecb}
\end{center}
\end{figure}

%%%%%%%%%%%%%%%%%% RECOMBINATION TEMP. & BALMER DECREMENT & REDDENING %%%%%%%%%%%%%%%%%%%%%%%%%%%%%%%%%%%%%%%%%

\subsection{Recombination temperatures and the Balmer decrement}\label{sec:final}

\subsubsection{Behavior of the recombination temperatures}\label{sec:rec}

The abundance discrepancy factor can be defined as the ratio, or the logarithmic difference, between abundances of a same ion derived from ORLs and CELs.
The ADF is a result of the large difference in the dependence of CELs and ORLs on temperature. To evaluate how the different models presented so far can account for the ADFs reported in the literature, we computed the recombination temperature characterizing each model. To reflect the fact that recombination line emissivities decrease with increasing temperature, we averaged the quantity $T_{rec}^{\alpha}$ for each selected ion X$^{+i}$ using a weight across the nebular structure that is proportional to the product of this ion density with the electron density. We therefore define the average recombination temperature of species X$^{+i}$ as $\trecx = \langle T_{rec}^{\alpha}\rangle^{1/\alpha}_{X+i}$.
We derived in this way the three recombination temperatures \trech, \treche\ and \treco, corresponding to the ions H$^{+}$, He$^{+}$ and O$^{++}$, respectively. We adopted $\alpha=-0.83$ (Eq.\,\ref{eq:trec}) for H, He  and $-0.87$ for O. Since in any model \trech\ and \treche\ do not differ by more than 30\,K, the temperature \treche\ will not be discussed further.

To characterize the behavior of \treco\ as a function of \toiii, we present various sequences of models (solid lines) in Figs.\,\ref{fig:treca} and \ref{fig:trecb}. The metallicity is the only parameter that varies along any of the sequences. A dotted line is used when plotting \trech, but it is only shown for cases where one model of the sequence has \trech\ differing from \treco\ by more than 3\%. The behavior of \treco\ in the case of $\kappa$--distribution (with $\kappa = 20$, 10 and 5) is shown in Fig.\,\ref{fig:treca} together with the case of \tsq\ models (with $\tsq=0.05$ or 0.10).  The $Z$ sequence for the combined shock-photoionization case with $\vs \simeq 50\,$\kms is overlaid as well. Interestingly, the \tsq\ sequences show a behavior similar to those of $\kappa$--distributions. Since $\tsq=0.10$ is close to the highest value reported by PG12, and since the corresponding sequence (purple line)  overlaps the $\kappa =10$ sequence, we infer that values as low as $\kappa \simeq 5$ should not be required to reproduce even the largest ADFs observed in some \hii\ regions. NDS12 came to the same conclusion using different temperature diagnostics.

All models representing metallicity inhomogeneities previously shown in Figs.\,\ref{fig:met} and \ref{fig:lmet} are displayed in Fig.\,\ref{fig:trecb} (see the figure line color legend for values of $Z_A$ and $\theta$). Wherever \treco\ substantially exceeds \toiii, the models imply a large ADF, which happens only with the higher metallicity sequences. A pertinent result from  Fig.\,\ref{fig:trecb} is that for $Z_A \la 0.05$ sequences, the models diverge very little from the equal temperature gray dashed line. Moreover, the upward curvature\footnote{The condition for a lower \treco\ than \toiii\ at intermediate weight values $0<w_i<1$ is that the relative line flux increase (or decrease) between models\,A and B must be higher for \oiiitw\ than for \oiiiw, that is: $\lg(F_{4363}^A/F_{4363}^B)/\lg(F_{5007}^A/F_{5007}^B) > 1$.} above the equal temperature line  (\ee{model sequences} nos.\,5--8 in Fig.\,\ref{fig:trecb}) is opposite to that of the higher $Z$ sequences, since \treco\ in \ee{model sequences nos.}\,5--8 lie above \toiii.  These models become unsuitable to account for any ADF among metal-poor nebulae. If we increase  the contrast $\theta$ much beyond the assumed value of 0.1\,dex, we find that it is possible to reverse the sign of the curvature. For instance, for the navy sequence with $Z_A \la 0.01$, the metallicity contrast $\theta$ had to be increased to 2.4\dex\ for \treco\ to become again lower than \toiii. Even though obtaining $\treco > \toiii$ is always possible with sufficiently high  $\theta$ values, targeting metal-poor nebulae  comes with the proviso that only low values of $w_i$ are allowed. It appears to us that such a fine-tuning in the parameter space is an implausible requirement. Unlike the alternate mechanisms presented in previous Fig.\,\ref{fig:treca}, metallicity inhomogeneities cannot account for sizable ADFs in nebulae that have $\toiii \ge 15\,000$\,K, that is, in \hii\ regions that are metal-poor. Yet, in their recalibration of the Pagel method, PG12 reported five \hii\ regions beyond $>15\,000$\,K. These nebulae are characterized by  an equivalent \tsq\ in the range 0.02--0.12, indicative of a temperature \treco\ markedly \emph{lower} than \toiii.

\subsubsection{Enhanced Balmer decrement and dereddening}\label{sec:bal}

\begin{figure}
\begin{center} \leavevmode
\includegraphics[width=0.49\textwidth]{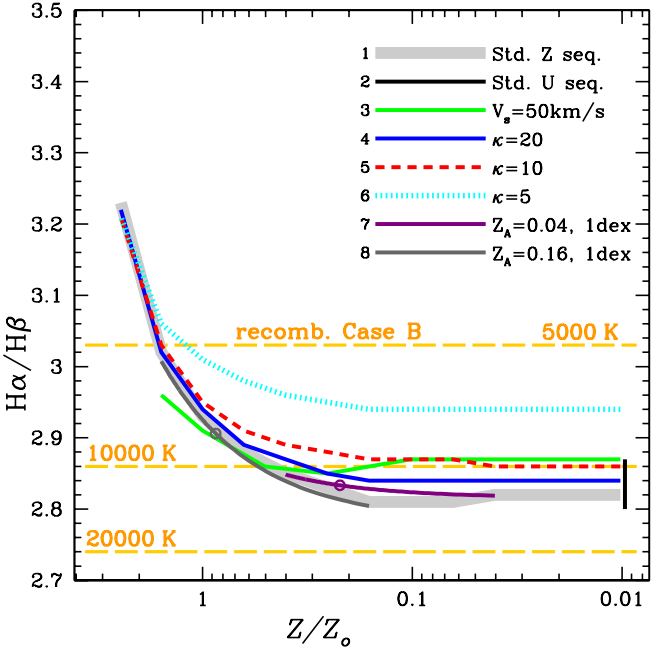}
\caption{Calculations of the Balmer decrement as a function of metallicity for various models. The thick light gray line is a metallicity sequence representing the homogeneous pure photoionization case with $\up=0.01$ (same sequence as in Fig.\,\ref{fig:figh1}). The (short vertical) black line shows a \up--sequence at a fixed metallicity $Z=0.01\,$\zsol\ (see cyan dotted line in Fig.\,\ref{fig:figh1}). Three metallicity sequences representing the $\kappa$--distribution case are shown. {The corresponding $\kappa$ value appears in the line color legend.} The light green line represents a metallicity sequence for a 50\,\kms\ shock combined with photoionization. Two sequences (labeled\,7 and 8 in the legend) are shown that represent models with metallicity inhomogeneities. The values of $Z_A$ (in \zsol\ units) and $\theta$ (in dex) appear in the line color legend. An open circle indicates the position where $w_i=0.5$. Yellow dashed lines represent the low-density recombination case-B Balmer decrement at three different temperatures (Osterbrock 1989). }
\label{fig:bal}
\end{center}
\end{figure}

The reddening correction that is routinely applied to observed line intensities can be a concern, even if the value of the Balmer decrement adopted for determining \av\ is apparently consistent with the subsequently derived \oiii\ temperature. In particular, an excessive dereddening  of  the auroral and nebular line intensities would result in overestimates of the \toiii\ and other CEL temperatures when these auroral lines occur at shorter wavelengths than the corresponding nebular lines\footnote{This is the case for \nii, \oiii, \siii\ and \ariii, but \emph{not} for \oii\ and \sii.}.  This inconsistency can have  consequences for the case of $\kappa$--distributions or large temperature inhomogeneities, since the intrinsic Balmer decrement is not set by \toiii, but rather arises from a plasma at a considerably lower temperature, \trec.  For instance, the intrinsic Balmer decrement can approach values $\simeq 3$ when the abundances are above solar and $\trec \ll \toiii$. A better understanding of temperatures in \hii\ regions may require that we relate the process of dereddening with temperature determination. To illustrate our concern, we carried out a comparison of the Balmer decrement from models that reproduce the spread in \toiii. The results are shown\footnote{A metallicity sequence with $\tsq=0.1$ was calculated, but it is not shown for clarity in Fig.\,\ref{fig:bal}.  The \tsq\ sequence closely follows the thick light gray line.} in Fig.\,\ref{fig:bal} where the calculated Balmer decrement is plotted as a function of nebular metallicity. Two metallicity inhomogeneity sequences with $\theta=1\,$dex are plotted as a function of average  metallicity defined as $\bar Z_i= [w_i+(1-w_i)10^{\theta}] Z_A$. These  correspond to a model\,A with $Z_A=0.04$\,\zsol\ (purple line) and  $Z_A=0.16$\,\zsol\  (dark gray line), respectively.  We note that they do not depart appreciably from the homogeneous photoionization case. The combined photoionization-shock models show evidence at \emph{low} metallicities of a slightly higher decrement. For $\kappa$--distributions, the lower the adopted $\kappa$ value, the higher the decrement toward \emph{low} metallicities.

We recall that the lower left quadrant in our temperature diagram contains objects whose metallicities are likely to be higher than 0.2\,\zsol\ (see Sect.\,\ref{sec:cal}). Yet, for our data set, the objects with the lowest \tsiii\ values can have corresponding \toiii\ temperatures that exceed  10\,000\,K. At metallicities $\sim 1$--2\,\zsol, which $\kappa$--distribution models favor to reproduce the strongest excesses in \toiii\ (red line in Fig.\,\ref{fig:figh7}), the Balmer decrement predicted by these models in Fig.\,\ref{fig:bal} is nearing or even exceeding 3. Hence, when comparing data with models that predict different  behaviors of CEL temperatures, it would make sense to first apply a reddening correction consistent with the Balmer decrement that these same models predict.

%It is apparent that assuming a  Balmer decrement around 10\,000\,K just because \toiii\ lies around that value would lead in the case of solar metallicity nebulae to an overestimate of the reddening in the event of a departure from the Maxwell-Boltzmann energy distribution, which in turn would lead to the measurement of excessive CEL temperatures and  undervalued nebular abundances.

%%%% CONCLUSIONS    %%%%%%%%%%%%%%%%%%%%%%%%%%%%%%%%%%%%%%%%

\section{Conclusions} \label{sec:con}

An analysis of 102 pairs of \oiii\  and \siii\ temperatures of \hii\ galaxies, giant extragalactic \hii\ regions, Galactic \hii\ regions, and \hii\ regions from the  Magellanic Clouds was carried out. The \oiii\ temperatures appear to be higher than the corresponding values derived from \siii\ in objects with \tsiii\ lower than 14\,000\,K. This trend is present in  objects with metallicities $Z \ga 0.2$. For the coolest nebulae (the highest metallicities) the \oiii\ temperature excess can reach  +3\,000\,K.  The results from grids of models that attempt to account for this excess can be summarized as follows:
   \begin{enumerate}
      \item Simple photoionization calculations result in essentially equal \oiii\ and \siii\ temperatures in the range of interest ($\tsiii<14\,000\,$K).  Varying either the ionization parameter \up, the metallicity $Z$, or the hardness of the ionizing \sed\ over a wide range does not alter this result.
      \item Including temperature inhomogeneities of large mean squared amplitude (\tsq) does not result in values of \toiii\ higher than \tsiii\ in the quadrant of reliable data ($\tsiii<14\,000\,$K).
      \item Metallicity inhomogeneities can successfully reproduce the observed excess in \toiii\ temperatures. By combining two photoionization models of widely differing metallicities (for instance a model with $Z=0.2\,\zsol$ with another with $Z=1.26\,\zsol$), most of the observed spread in temperatures can be reproduced.
      \item Adopting a $\kappa$--distribution for the electron energy distribution results in models that successfully reproduce the observed excess in \toiii\ temperatures. Values of $\kappa$ in the range $\ga 5$ to 20 are favored.
      \item Shock waves that propagate in the photoionized gas can also account for the observed excess in \toiii. The 1D  models that we computed are simplistic and cannot be considered a fully autoconsistent description of shock waves propagating in a chaotic medium. 3D hydrodynamical calculations would be required that include proper radiative transfer in the context of outward-propagating shocks.
      \item Both the shocked nebula model and the photoionization model with abundance inhomogeneities share the property of a diminishing excess in \toiii\ with decreasing nebular metallicity, while the $\kappa$--distributions result in a near constant \toiii--\tsiii\ offset when $\kappa$ is set to a fixed value. Through improved observations, it would be important to determine the behavior of the temperatures in the top right quadrant of the \tsiii\ vs. \toiii\ diagram.
      \item We derived the recombination temperatures for the various models. Models that consider $\kappa$--distributions or temperature inhomogeneities \tsq\ result in recombination temperatures that are substantially lower than \toiii\ over the whole metallicity range. On the other hand, models with metallicity inhomogeneities have \treco\ \emph{higher} than \toiii\ in the upper right quadrant where $\toiii\ga 15\,000$\,K. Metallicity inhomogeneities are therefore only applicable to metallicity-enriched nebulae and could not account for any sizable ADF in metal-poor nebulae ($Z<0.2$\,\zsol).
   \end{enumerate}

\begin{acknowledgements}
      LB acknowledges support from \emph{CONACyT} grant CB-128556. RM received financial support from project IN105511 from PAPIIT (UNAM). We are thankful to Jana Benda Klouda for providing us with the initial translation in English. We thank Diethild Starkmeth for proofreading the article. We acknowledge that many suggestions from an anonymous referee contributed to improve the usefulness and scope of the final version.
\end{acknowledgements}

%%%%%%%%%%%%%%%%%%%%%%%%%   BIBLIOGRAPHY  %%%%%%%%%%%%%%%%%%%%%%%%%%%%%%%%%%%%%%%%%%%%%%

%%% Faltabann las siguientes referencias: %%%%%%%%%%%%%%%%%%%%%%%%%%%%%%%%%%%%%%%%%%%%%%%%
%%% Now OK   Shields 1974
%%% Now OK   año 1998, no 1999 Kingdom & Ferland 1998
%%% Now OK   Hagele et al. 2010
%%% Now OK   Definir en el texto BLOO (Binette et al. 2000)
%%% DONE     Mencionar en el texto Raga & Riera 2008    y    Viegas 97   y    Aldrovandi

%%%%%%%%%%%%%%%%%%%%%   APPENDIX IN PAPER  %%%%%%%%%%%%%%%%%%%%  APPENDIX  ********
\appendix
\section{Sample abundance calibration}\label{sec:app}

\begin{figure}[ht]
\begin{center} \leavevmode
\includegraphics[width=0.49\textwidth]{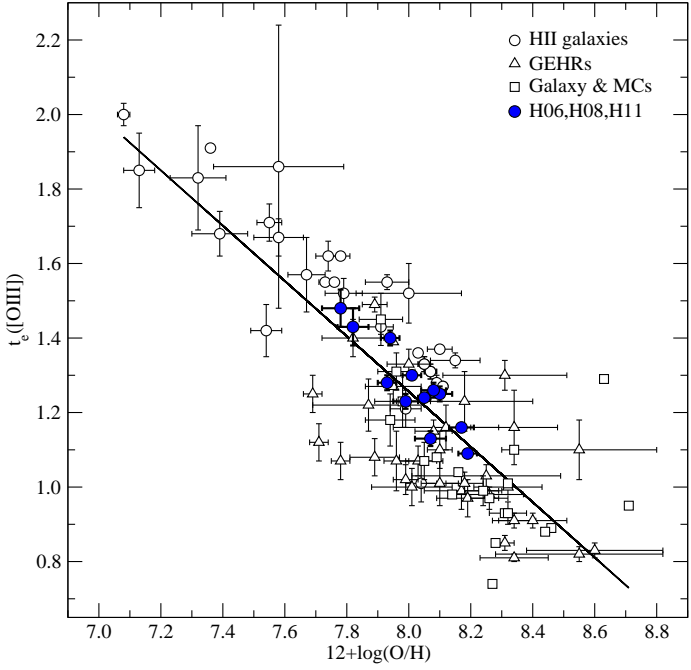}
\caption{Relation between t$_e$(\oiii) and total oxygen abundances ($12+\lg({\rm O/H})$) for
\hii\ galaxies from H06, H08 and H11 (solid circles), and \hii\
galaxies (open circles), GEHRs (upward triangles) and diffuse \hii\ regions in
the Galaxy and the Magellanic Clouds (squares) from P\'erez-Montero \& D\'iaz
(2005). The solid line is the actual fit to the data. The temperatures are in
units of 10$^4$\,K. }
\label{fig:oxy}
\end{center}
\end{figure}

Figure~\ref{fig:oxy} shows the relationship between the electron temperatures
inferred from the \oiii\ lines and the total oxygen abundances $12+\lg$(O/H). This sample was built from the literature and is a collection of objects for which measurements of the nebular and auroral emission lines of \oii\ and \oiii\ were available. This sample comprises objects from P\'erez-Montero \& D\'iaz (2005) for which the determination of \toiii\ and $12+\lg$(O/H) was possible, using the direct method (see H\"agele et\,al. 2008: H08), as well as the \hii\ galaxies from H06, H08, and H11. The solid line corresponds to a least-squares fitting of the data, taking their individual
errors into account:
%\[t_e([O\textrm{\sc iii}])\,=\,(0.74\,\pm\,0.05)\,(12+\lg(O/H))\,+\,(7.2\,\pm\,0.4)\] missing minus sign
$$t_e([\oiii])=-0.74\pm{0.05} \times \{12+\lg({\rm {O/H}})\} + 7.2\pm{0.4} \; ,$$
\noindent where t$_e$(\oiii) is the \oiii\ electron temperature in units of 10$^4$\,K. The O/H determination above is based on CELs only. It likely underestimates the true O/H abundance ratio by a factor $\sim 2$, which would likely have been inferred if the ADF had been determined from ORLs and had been taken into account in the above regression. If the electron energies exhibit a $\kappa$ non-equilibrium distribution, this is expected to increase the oxygen abundance, as shown by NDS12. Likewise with models that consider temperature or abundance inhomogeneities.

\section{Behavior of combined shock-photoionization models when \vs\ is increased}\label{app:sho}

In combined shock-photoionization models the  span in temperature and in ionization stages of metals covers a much wider range than with photoionization alone. Furthermore, the temporal evolution of the plasma that characterizes shocks is another significant difference.
To analyze why the \toiii\ temperatures increase markedly with \vs, we compare two models from the cyan sequence with $Z=1.0\,$\zsol\ in Table\,\ref{tab:sho}. When one derives volume-averaged quantities that are weighted by the density square of the ion O$^{+2}$ or S$^{+2}$, it is found that the 45.3\,\kms\ and 56.1\,\kms\ models are characterized by a volume-averaged ionic temperature $\left<{T_{O++}}\right>$ of 8\,400\,K and 9\,700\,K, respectively, while for $\left<{T_{S++}}\right>$  the corresponding values are 8\,070\,K and 7\,980\,K, respectively. Hence, not much separates the two models as far as average properties are concerned.
In the 45.3\,\kms\ model, the \emph{mean} doubly ionized fraction for oxygen $\left<{n_{O++}/n_O}\right>$  is 0.164 while the equivalent fraction for sulphur $\left<{n_{S++}/n_S}\right>$ is 0.889. In the 56.1\,\kms\ model, the \emph{mean} doubly ionized fractions are 0.087 and 0.844, respectively. Photoionization is responsible for the small reverse effect of a reduction in $\left<{n_{O++}/n_O}\right>$ because the \emph{mean} density $\left<{n_{H+}}\right>$ actually increases from 21.1\,\cmc\ to 32.1\,\cmc, (between the 45.3\,\kms\ and the 56.1\,\kms\ model, respectively), which has the effect of reducing the effective ionization parameter. We can conclude that combined photoionization-shock models in this velocity regime do not drastically alter the \emph{mean} degree of ionization of oxygen or sulphur, or their mean temperatures where the doubly ionized ions reside. Indeed, both shock models start with the same ionization fraction at the shock front, of 0.93 for $n_{O++}/n_O$ and 0.49 for $n_{S++}/n_S$, but they possess quite different postshock temperatures (35\,000\,K and 50\,400\,K, respectively).

To summarize, there are two main factors that account for the behavior of \toiii\ in Fig.\,\ref{fig:sho}:  (1)  $\left<{n_{O++}/n_O}\right> \ll \left<{n_{S++}/n_S}\right>$, which causes the much hotter region near the shock front to contribute proportionally more to the integrated line spectrum of the \oiii\ lines than it does for the  \siii\ lines, and (2) the strong dependance of the CELs on the factor $\propto T^{-0.5} \exp(-\Delta E/kT)$  is responsible for favoring the copious emission of lines that have the highest energy transitions $\Delta E$ at the head of the shock where $T\la \tpost$, as is the case for \oiiitw. This is the main factor  that accounts for the high $4363\AA/5007\AA$ line ratio, and hence for the higher values of \toiii. \siiitw\ on the other hand does not beneficiate much from either factor.

The decrease of both \toiii\ and \tsiii\ at low velocities in sequences of very low-metallicity gas (light green line and red line sequences in Fig.\,\ref{fig:sho}) reflects the increase in the mean density as the postshock temperature is successively increased, which causes a reduction of the effective ionization parameter and hence an increase in the neutral fraction H$^0$/H, which favors an increase in the cooling due to  collisional excitation of \hi. This increased cooling in turn decreases $\left<{T_{O++}}\right>$ and $\left<{T_{S++}}\right>$. A similar mechanism is at work at very low metallicities in \up--sequences of homogeneous photoionization models (e.g. cyan dotted line in Fig.\,\ref{fig:figh1}), as discussed in Sect.\,\ref{sec:pure}.
%%%%%%%%%%%%%%% END  END   END  %%%%%%%%%%%%%%%%%%%%%%%%%%%%%%%%%%%%%%%%%%%%%%%%%%%%%%%%%%%%%%%%%%%%%%%%%%%%%%%

\end{document}